\documentclass{pasj01}
\usepackage{bm}

\usepackage{lineno}
\usepackage{color}

\newcommand{\mdotinaves}{380}
\newcommand{\mdotoutaves}{0.13}

\newcommand{\mdotoutends}{230}
\newcommand{\rqsss}{1000}

\newcommand{\rlaus}{1000}

\newcommand{\rlaucyls}{820}
\newcommand{\rlauinfcyls}{480}

\newcommand{\mdotinavef}{140}

\newcommand{\rqssf}{400}

\Received{$\langle$1-Jun-2022$\rangle$}
\Accepted{$\langle$1-Sep-2022$\rangle$}

\begin{document}

\title{ 
  ~Large-scale outflow structure and radiation properties of super-Eddington flow:                             
  Dependence on the accretion rates   
}

\author{Shogo \textsc{YOSHIOKA}\altaffilmark{1}}
\altaffiltext{1}{Department of Astronomy, Graduate School of Science, Kyoto University, Kitashirakawa-Oiwake-cho, Sakyo-ku, Kyoto 606-8502, Japan}
\email{yoshioka@kusastro.kyoto-u.ac.jp}

\author{Shin \textsc{Mineshige}\altaffilmark{1}}

\author{Ken \textsc{Ohsuga}\altaffilmark{2}}
\altaffiltext{2}{Center for Computational Sciences, University of Tsukuba, Ten-nodai, 1-1-1 Tsukuba, Ibaraki 305-8577, Japan}

\author{Tomohisa \textsc{Kawashima}\altaffilmark{3}}
\altaffiltext{3}{Institute for Cosmic Ray Research, The University of Tokyo, 5-1-5 Kashiwanoha, Kashiwa, Chiba 277-8582, Japan}

\author{Takaaki \textsc{Kitaki}}



\maketitle

\begin{abstract}
  In order to precisely evaluate the impacts by super-Eddington accretors to their environments,
  it is essential to assure a large enough simulation box and long computational time to avoid any artefacts from numerical settings as much as possible. 
  In this paper, we carry out axisymmetric two-dimensional radiation hydrodynamic simulations
  around a $10~M_\odot$ black hole in large simulation boxes
  and study the large-scale outflow structure and radiation properties of 
  super-Eddington accretion flow for a variety of black hole accretion rates, 
  ${\dot M}_{\rm BH} = (110 – 380) ~L_{\rm Edd}/c^2$ (with $L_{\rm Edd}$ being 
  the Eddington luminosity and $c$ being the speed of light).
  The Keplerian radius of the inflow material, at which centrifugal force balances with gravitational force, 
  is fixed to 2430 Schwarzschild radii. 
  We find that the mechanical luminosity grows more rapidly than 
  the radiation luminosity with an increase of ${\dot M}_{\rm BH}$.
  When seen from a nearly face-on direction, especially, 
  the isotropic mechanical luminosity grows in proportion to ${\dot M}_{\rm BH}^{2.7}$,
  while the total mechanical luminosity is proportional to ${\dot M}_{\rm BH}^{1.7}$.
  The reason for the former is that 
  the higher ${\dot M}_{\rm BH}$ is, the more vertically inflated becomes the disk surface,
  which makes radiation fields more confined in the region around the rotation axis, thereby strongly accelerating outflowing gas.  
  The outflow is classified into pure outflow and failed outflow, depending whether
  outflowing gas can reach the outer boundary of the simulation box or not.
  The fraction of the failed outflow decreases with a decrease of ${\dot M}_{\rm BH}$.
  We analyze physical quantities along each outflow trajectory, 
  finding that the Bernoulli parameter ($Be$) is not a good indicator to discriminate pure and failed outflows, 
  since it is never constant because of continuous acceleration by radiation-pressure force.
  Pure outflow can arise, even if $Be < 0$ at the launching point.

\end{abstract}

\section{Introduction}
\label{sec-introduction}
It is well known that accretion disks around compact objects, 
such as black holes and neutron stars, 
can very efficiently release gravitational potential energy (see, e.g., Chap. 1 of Kato et al. 2008 for a review). 
The released gravitational energy is converted into radiation energy and/or mechanical energy of the gas.
There is a classical limit to the total amount of radiation energy released per unit time by accretion;
that is what we call the Eddington luminosity $L_{\rm Edd}.$
It is defined by the balance between radiation force and gravitation force in the spherically symmetric accretion assumption,
and is written as,
\begin{eqnarray}
  L_{\rm Edd}&\equiv&\frac{4\pi cGM_{\rm BH}}{\kappa_{\rm es}}\simeq 1.26\times10^{39}\left(\frac{M_{\rm BH}}{10M_{\odot}}\right)~{\rm erg}~{\rm s^{-1}}.
\end{eqnarray}
Here, $c$ is the light speed, $G$ is the gravitational constant, $M_{\rm BH}$ is the mass of a central black hole, $\kappa_{\rm es}$ is the Thomson scattering opacity,
and we assumed pure hydrogen abundance. 
It is now widely accepted that the classical limit can be exceeded in disk accretion 
because of the separation of the directions of gas inflow and of the radiation output.
The accretion flow with an accretion rate $\dot{M}_{\rm BH}$ much greater than $L_{\rm Edd}/c^2$ is called a super-Eddington accretion
(see, e.g., Chap. 10 of Kato et al. 2008 for a review).

Among diverse black hole objects, super-bright compact sources called Ultra-luminous X-ray sources (ULXs) exhibit rather unique observational features;
they are bright, with luminosities being over $10^{39}~ {\rm erg\ s^{-1}}$,
but they are located at a off-center regions; that is, they are not active galactic nucleus (AGNs) (Long et al. 1981; Fabbiano 1989; Soria et al. 2007; Kaaret et al. 2017).
Their central engines are still under discussion; promising models include
(1) super-Eddington accretion onto a stellar mass black hole (Watarai et al. 2001; King et al. 2001; Gladstone et al. 2009; Sutton et al. 2013; Kawashima et al. 2012; Motch et al. 2013; Middleton et al. 2015; Kitaki et al. 2017),
(2) super-Eddington accretion onto a neutron star (Basko \& Sunyaev 1976; Mushtukov et al. 2018; Bachetti et al 2014; F\"{u}rst et al. 2016; Kawashima et al. 2016; Israel et al. 2017a; Takahashi et al. 2018; Carpano et al. 2018),
and (3) sub-Eddington accretion onto an intermediate-mass black hole (Makishima et al. 2000; Miller et al. 2004; Strohmayer \& Mushotzky 2009; Miyawaki et al. 2009).
Here, we focus our discussion on the first case.

The super-Eddington accretion flow has two key signatures: 
it can shine in excess of the Eddington luminosity, and it has powerful outflows due to the increase in radiation force (Ohsuga et al. 2005; Takeuchi et al. 2010, Poutanen et al. 2007).
In particular, outflow is crucially important, since it carries mass, momentum, and energy of gas
to the surrounding environment and can assert a significant impact there (Regan et al. 2018; Takeo et al. 2020; Hu et al. 2022; Botella et al. 2022).

To understand the nature of super-Eddington accretors, 
it is essential to solve the interaction between the radiation and the gas;
that is, the radiation hydrodynamics (RHD) simulations are necessary (Eggum et al. 1987, Fujita \& Okuda 1998, Ohsuga et al. 2005; Narayan et al. 2017; Ogawa et al. 2017; Takeo et al. 2018; Kitaki et al. 2018). 
Such RHD simulations have been extensively performed in these days,
followed by radiation magnetohydrodynamics (RMHD) simulations (e.g., Ohsuga et al. 2009, 2011; Jiang et al. 2014, 2019).
Furthermore, some of the RHD/RMHD simulations are under the general relativistic (GR) formalism (McKinney et al. 2014, 2015, 2017;  Fragile et al. 2014; S\c{a}dowski et al. 2015, 2016; Takahashi et al. 2016).

Here, we wish to point out two key issues involved with most of the current RHD/RMHD simulation studies:
\begin{enumerate}
\item Small box size. The size of the computational box is limited due to the restriction from the computer side. This could lead to overestimation of outflow rate (explained later).
\item Small angular momenta of injected gas. 
It is thus difficult to investigate the case of the ULXs, in which 
the injected materials, presumably supplied from the companion star, seem to have relatively large angular momenta.
\end{enumerate}

It will be useful to define the two key radii; 
(1) the Keplerian radius, $r_{\rm K}$,  at which the centrifugal force balances with the gravitational force
for a given specific angular momentum of the injected gas,
and (2) the photon trapping radius, $r_{\rm trap}$, inside which photon trapping is effective (Begelman 1978; Ohsuga et al. 2005).
If we assume a small Keplerian radius, $r_{\rm K} < r_{\rm trap}$,
injected material accumulates inside the trapping radius, forming a puffed-up region,
from which significant outflow emerges. 
This may lead to overestimation of
the outflow rate $\dot{M}_{\rm outflow}$ (Kitaki et al. 2021; hereafter K21). 
If we take small computational boxes, moreover,
some of outflow that falls back into the disk after launch (failed outflow) could be
mis-classified as a pure outflow that successfully escape from the system.
This will also lead to overestimation of outflow rates.

To avoid such numerical artefacts as much as possible, K21 performed two-dimensional (2D) axisymmetric RHD simulations, 
assuming (1) a large Keplerian radius, $r_{\rm K}=2430~r_{\rm S}$, and
adopting (2) a large simulation box of a size of $r_{\rm out}=3000~r_{\rm S}$
so that they could elucidate the disk-outflow structure over a wide region across $r_{\rm trap}$.
Their simulation was, however, restricted to only one parameter-set case.
We wish to expand parameter space to get more general view of super-Eddington outflow.
This is the primary aim of the present study.

We perform the same type of axisymmetric 2D-RHD simulations as that of K21 but for a variety of mass accretion rates under realistic simulation settings.
The key questions that we address in the present study are two-fold:
(Q1) How do the radiation and mechanical luminosities depend on $\dot{M}_{\rm BH}$ and viewing angle, 
and (Q2) how much material is launched from which radii and to which directions?
The plan of the present paper is as follows: We explain calculated models and numerical methods in section 2, and present our results of large-scale outflow structure in section 3.
There, we emphasize the $\dot{M}_{\rm BH}$ dependence of the radiation and outflow properties.
We then give discussion in section 4. 
The main issues to be discussed are the impact on the environments, energy conversion efficiency, 
connection with the observations of ULXs, and the Bernoulli parameter along the streamline.
The final section is devoted to concluding remarks.

\section{Calculated Models and Numerical Methods}
\subsection{Radiation hydrodynamics Simulations}
\label{sec-rhd}
In the present study, we consider super-Eddington accretion flow and associated outflow around
black hole with mass of 10 $M_\odot$.
We inject mass with a certain amount of angular momentum from the outer simulation boundary at a constant rate
(more quantitative description will be given later).
For calculating radiation flux and pressure tensors, we adopt the
flux-limited diffusion approximation (Lervermore \& Pormaraning 1981; Turner \& Stone 2001).
Since we do not solve the magnetic fields in the present simulation and thus
adopt the $\alpha$ viscosity prescription (Shakura \& Sunyaev 1973).
General relativistic effects are taken into account by employing the pseudo-Newtonian potential (Paczy\'{n}sky \& Wiita 1980).

Basic equations and numerical methods are the same as those in K21 (see also Ohsuga et al. 2005; Kawashima et al. 2009).
We solve the axisymmetric two-dimensional radiation hydrodynamics equations     
in the spherical coordinates $(x,y,z)=(r\sin\theta\cos\phi, r\sin\theta\sin\phi, r\cos\theta)$,
where the azimuthal angle $\phi$ is set to be constant and the $z$-axis coincides with the rotation-axis.
We put a black hole at the origin. 
In this paper, we distinguish $r$, radius in the spherical coordinates, and $R\equiv\sqrt{x^{2}+y^{2}}$, radius in the cylindrical coordinates. 

The basic equations are explicitly written as follows:
The continuity equation is
\begin{eqnarray}
  \frac{\partial \rho}{\partial t}+\nabla\cdot\left(\rho\bm{v}\right)=0,
\end{eqnarray}
where $\rho$ and $\bm{v}=(v_{r},v_{\theta},v_{\phi})$ 
is the gas mass density and the velocity of gas, respectively. 

The equations of motion are
\begin{eqnarray}
\frac{\partial(\rho v_{r})}{\partial t}+\nabla\cdot\left(\rho v_{r}\bm{v}\right)
= -\frac{\partial p}{\partial r}
+\rho\left[\frac{v_{\theta}^2}{r}+\frac{v_{\phi}^2}{r}
-\frac{GM_{\rm BH}}{(r-r_{\rm s})^2}\right] \nonumber\\
+\frac{\chi}{c}F_{0}^{r},  \label{eom_r}\\
\frac{\partial (\rho rv_{\theta})}{\partial t}+\nabla\cdot\left(\rho rv_{\theta}\bm{v}\right)
= -\frac{\partial p}{\partial \theta}+\rho v_{\phi}^{2}\cot\theta 
+r\frac{\chi}{c}F_{0}^{\theta}, \label{eom_th}\\
\frac{\partial(\rho r\sin\theta v_{\phi})}{\partial t}+\nabla\cdot(\rho r\sin\theta v_{\phi}\bm{v})
= \frac{1}{r^{2}}\frac{\partial}{\partial r}\left(r^{3}\sin\theta t_{r\phi}\right),
\end{eqnarray}
where $p$ is the gas pressure,
$\chi=\kappa+\rho \sigma_{\rm T}/m_{\rm p}$ is the total opacity,
(with $\kappa$ being free-free and free-bound absorption opacity and $\sigma_{\rm T}$ 
being the cross-section of Thomson scattering, see Rybicki \& Lightman 1979),
$m_{\rm p}$ is the proton mass, and
$\bm{F}_{0}=(F_{0}^r,F_{0}^\theta,F_{0}^{\phi})$ is the radiative flux in the comoving frame. 
In the viscous-shear stress, only the $r$-$\phi$ component is assumed to be nonzero and is prescribed as
\begin{eqnarray}
  t_{r\phi}&=&\eta r \frac{\partial }{\partial r}\left(\frac{v_{\phi}}{r} \right),
\end{eqnarray}
with the dynamical viscous coefficient being
\begin{eqnarray}
  \eta&=&\alpha \frac{p+\lambda E_{0}}{\Omega_{\rm K}}.
  \label{dvis}
\end{eqnarray}
where $\alpha=0.1$ is the viscosity parameter,
$\Omega_{\rm K}$ is the Keplerian angular speed,
$E_{0}$ is the radiation energy density in the comoving frame,
and $\lambda$ represents the flux limiter of the flux-limited diffusion approximation.
We adopted the functional form, equation ({\ref{dvis}}), assuming that 
$\eta$ is proportional to the total pressure in the optically thick limit (since then we have $\lambda=1/3$ ), 
and that $\eta$ is proportional to the gas pressure in the optically thin limit (since then we find $\lambda = 0$) 
so that the adopted prescription should agree with that employed in the standard disk theory. 
Note that we have adopted the same prescription in the previous simulation studies 
(e.g., Ohsuga et al. 2005, Kawashima et al. 2009, Kitaki et al. 2021).
In practice, local radiation MHD simulations demonstrate that 
the magnitude of the effective alpha viscosity is proportional to the total pressure 
(i.e., radiation plus gas pressure) rather than the gas pressure only in the limit of optically thick, 
radiation pressure dominant disks (e.g., Turner et al. 2003).

The energy equation for gas is
\begin{eqnarray}
  \frac{\partial e}{\partial t}+\nabla\cdot\left(e\bm{v}\right)&=&-p\nabla\cdot\bm{v}-4\pi\kappa B+c\kappa E_{0}\nonumber\\
  &&+\Phi_{\rm vis}-\Gamma_{\rm Comp},
\end{eqnarray}
while the energy equation for radiation is
\begin{eqnarray}
  \frac{\partial E_{0}}{\partial t}+\nabla\cdot\left(E_{0}\bm{v}\right)&=&-\nabla\cdot\bm{F}_{0}-\nabla\bm{v}:\bm{{\rm P}}_{0}+4\pi\kappa B-c\kappa E_{0}\nonumber\\
  &&+\Gamma_{\rm Comp}.
\end{eqnarray}
Here $e$ is the internal energy density,
which is linked to the gas pressure through the ideal gas equation of state,
$p=(\gamma -1)e=\rho k_{\rm B}T_{\rm gas}/(\mu m_{\rm p})$
(with $\gamma=5/3$ being the specific heat ratio,
$k_{\rm B}$ being the Boltzmann constant,
$\mu=0.5$ being the mean molecular weight (we assume pure hydrogen plasmas),
and $T_{\rm gas}$ being the gas temperature, respectively),
$B=\sigma_{\rm SB}T_{\rm gas}^{4}/\pi$ is the blackbody intensity
(with $\sigma_{\rm SB}$ being the Stefan--Boltzmann constant),
$\bm{{\rm P}}_{0}$ is the radiation pressure tensor in the comoving frame, and
$\Phi_{\rm vis}$ is the viscous dissipative function;
\begin{eqnarray}
  \Phi_{\rm vis}=\eta\left[r\frac{\partial }{\partial r}\left(\frac{v_{\phi}}{r} \right)\right]^{2}.
\end{eqnarray}
The Compton cooling/heating rate 
is described as
\begin{eqnarray}
  \Gamma_{\rm Comp}&=&4\sigma_{\rm T}c\frac{k_{\rm B}\left(T_{\rm gas}-T_{\rm rad}\right)}{m_{\rm e}c^{2}}\left(\frac{\rho}{m_{\rm p}} \right)E_{0}.
\end{eqnarray}
Here, $m_{\rm e}$ is the electron mass and
$T_{\rm rad}\equiv (E_{0}/a)^{1/4}$ is the radiation temperature 
with $a=4\sigma_{\rm SB}/c$ being the radiation constant. 
Under the FLD approximation, $F_0$ and $\bm{{\rm P}}_{0}$ are calculated in terms of $E_0$. 
Because of the axisymmetry, FLD approximation gives $F_{0}^{\phi}=0$ in the whole calculation region.

  We wish to stress that Compton cooling/heating works not only in the optically thin disk atmosphere (in which $T_{\rm gas} > T_{\rm rad}$) 
  but also in the optically thick disk (in which $T_{\rm gas} \sim T_{\rm rad}$). 
  To prove if this is the case, we numerically checked the heating and cooling timescales
  at $r=200~r_{\rm S}$. 
  In the equatorial region, the timescale of viscous heating is comparable to that of Compton cooling, 
  whereas the timescale of bremsstrahlung cooling is longer than the other two by one order of magnitude or more. 
  We wish to note that $T_{\rm gas} \sim T_{\rm rad}$ in the disk region does not necessarily mean that 
  the Compton cooling/heating is unimportant, but rather mean that 
  $T_{\rm gas} \sim T_{\rm rad}$ is achieved as the result of efficient Compton cooling/heating.

  K21 investigated the magnitude of each term on the right-hand side of the gas energy equation (8) 
  and have concluded that the gas is heated by the viscous heating generated in the disk,
  but it is immediately converted to radiation energy through Compton scattering, 
  resulting in energy transport in the form of advection cooling.

\subsection{Initial conditions and calculated models}
\label{sec-initial-condition}

\begin{table*}[]
  \tbl{Model parameters}{
  \begin{tabular}{lll}
      \hline
      parameter & symbol & value(s) \\
      \hline \hline
      black hole mass & $M_{\rm BH}$  $ [M_{\odot}]$ & 10    \\
      mass injection rate & ${\dot M}_{\rm input}$ $ [L_{\rm Edd}/c^2]$ & 350, 500,  700,  2000 \\
      viscosity parameter & $\alpha$ & 0.1 \\
      simulation box: inner radius & $r_{\rm in}$ $[r_{\rm S}]$ & 2.0 \\
      simulation box: outer radius$^*$ & $r_{\rm out}$ $[r_{\rm S}]$ & 3000 or 6000 \\
      Keplerian radius & $r_{\rm K}$ $[r_{\rm S}]$ & 2430 \\ \hline
  \end{tabular}}
  \begin{tabnote}
    $^*$ We assign $r_{\rm out} = 6000 ~r_{\rm S}$ except in Model-180 (taken from K21), 
    in which we assigned $r_{\rm out} = 3000 ~r_{\rm S}$ (see also table 2 for calculated models).

  \end{tabnote}
  \label{parameter}
\end{table*}

As was already mentioned, we adopt a large Keplerian radius ($r_{\rm K}= 2430~r_{\rm S}$)
and large computational box size $r_{\rm in}=2~r_{\rm S}\leq r \leq r_{\rm out}=6000~r_{\rm S}$
(except one case, described later).
We only solve the upper-half domain above the equatorial plane; i.e., $0 \leq \theta \leq\pi/2$.

Grid points are uniformly distributed in logarithm in the radial direction;
$\triangle\log_{10} r = (\log_{10} r_{\rm out}-\log_{10} r_{\rm in})/N_{r}$,
while it is uniformly distributed in $\cos\theta$ in the polar direction;
$\triangle\cos \theta=1/N_{\theta}$, where the numbers of grid points     
are $(N_{r},N_{\theta})=(200,240)$ throughout the present study.

We initially put a hot optically thin atmosphere with negligible mass around the black hole for numerical reasons.
The initial atmosphere is assumed to be in uniform temperature distribution and hydrostatic equilibrium in the radial direction.
Then, the density profile is
\begin{eqnarray}
  \rho_{\rm atm}(r,\theta)&\equiv&\rho_{\rm out}\exp\left[\frac{\mu m_{\rm p}GM_{\rm BH}}{k_{\rm B}T_{\rm atm} r_{\rm out}}\left(\frac{r_{\rm out}}{r}-1\right)\right].
\end{eqnarray}
where $\rho_{\rm out}$ is the density at the outer boundary and $T_{\rm atm}$ is the temperature of hot optically thin atmosphere.
We employ $\rho_{\rm out}=10^{-17}{\rm g~cm^{-3}}$ and $T_{\rm atm}=10^{11}{\rm K}$, following Ohsuga et al. (2005).

Since the main purpose of this study is to investigate the ${\dot M}_{\rm BH}$ dependence 
of the super-Eddington flow and outflow, we fix the black hole mass and $\alpha$ viscosity parameter, 
while we vary mass injection rate $\dot{M}_{\rm input}$ (see table \ref{parameter}).

Matter is injected continuously at a constant rate of $\dot{M}_{\rm input}$
through the outer disk boundary at $r=r_{\rm out}$ and $0.49\pi\leq\theta\leq0.5\pi$.
We adopted a relatively smaller solid angle, but this does not necessarily mean a higher velocity for a fixed mass injection rate.
This is because although we assume the standard disk relations to determine the density and velocity of the injected gas for a given mass injection rate, 
the in-fall motion of the gas is soon accelerated to approach the free fall velocity because of small centrifugal force (note $r_{\rm K} \ll r_{\rm out})$. 
The injected gas is assumed to possess an specific angular momentum corresponding to the Keplerian radius of $r_{\rm K}=2430~r_{\rm S}$
(i.e., the initial specific angular momentum is $\sqrt{GM_{\rm BH} r_{\rm K}}$).
We thus expect that inflow material first falls towards the center
and forms a rotating gaseous ring with a radius of around $r\sim r_{\rm K}$, 
from which the material slowly accretes inward via viscous diffusion process.
We assume that matter freely goes out but not come in through the outer boundary ($r=r_{\rm out}$, $\theta$=$0 - 0.49\pi$) and the inner boundary ($r=r_{\rm in}$).

We assume that 
the density, gas pressure, radial velocity, and radiation energy density are symmetric at the rotation axis,
while $v_{\theta}$ and $v_{\phi}$ are the antisymmetric.
On the equatorial plane, on the other hand,
$\rho$, $p$, $v_{r}$, $v_{\phi}$ and $E_{0}$ are symmetric,
and $v_{\theta}$ is antisymmetric.
See Ohsuga et al. (2005) for more detailed descriptions regarding the boundary conditions.

\section{Results: Large-Scale Outflow Structure}
\subsection{Overall flow structure}
\label{sec-overall}

\begin{figure*}[]
  \begin{center}
    \includegraphics[width=145mm, bb= 0 0 697 568]{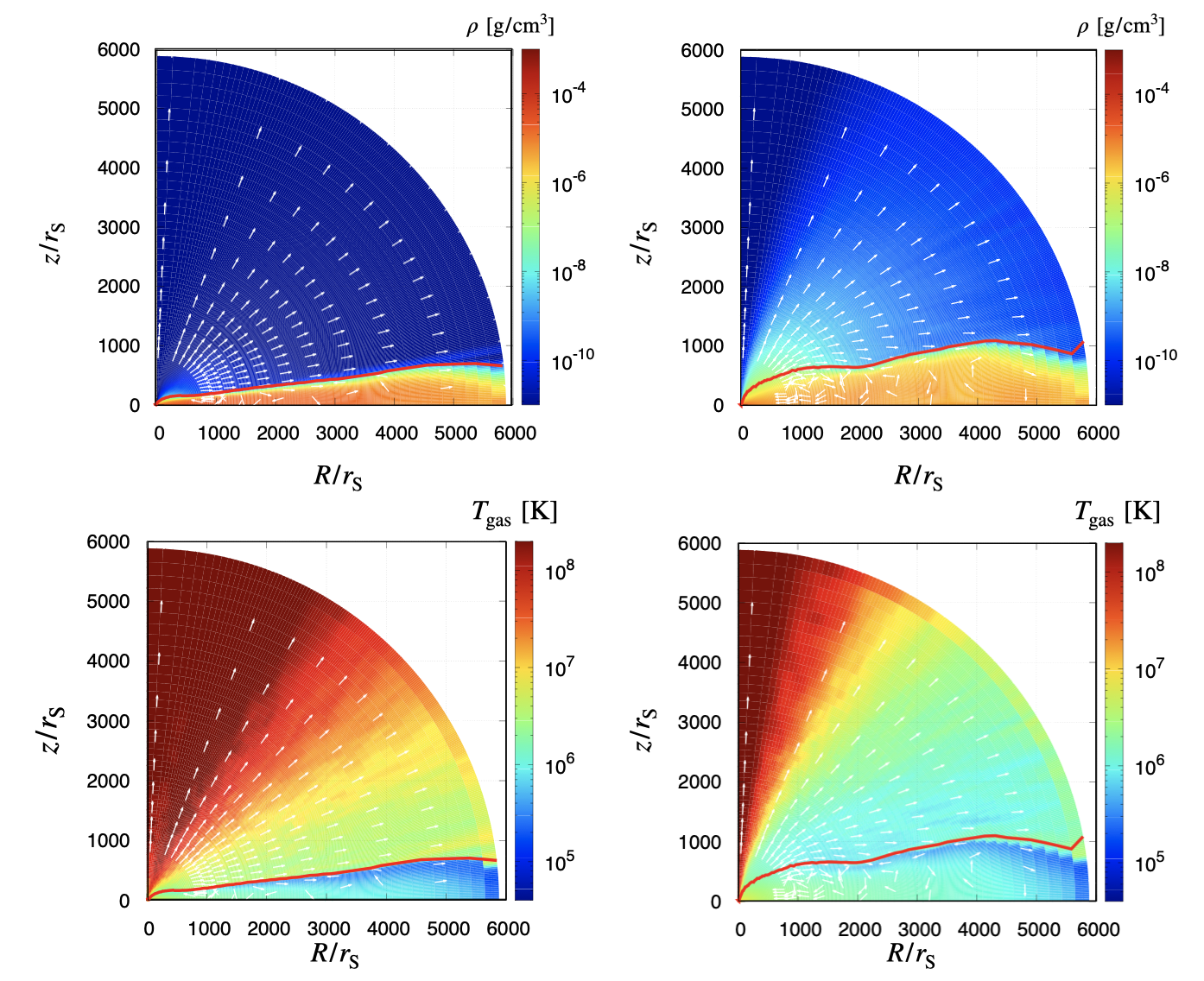}
  \end{center}
  \caption{
    Time-averaged density distributions (upper) and gas temperature distributions (lower) of super-Eddington accretion flow 
    and associated outflow around a black hole in Model-140 (left) and Model-380 (right), respectively.
    Time average was made during the interval of $t\sim 23800$ -- $24300~{\rm sec}$.
    Overlaid are the gas velocity vectors whose lengths are proportional to the logarithm of the absolute velocity.
    The red line represents the disk surface, which is defined as the loci where the radiation force balances the gravitational force.
  }
  \label{fig1}
\end{figure*}

In this paper, we examine the large-scale time-averaged structure of inflow and outflow in a quasi-steady state, unless stated otherwise.
We ran the simulation for 0 -- 24500 sec.
 (See figure \ref{figa1}  in Appendix A for the light curves of some models.)
We first show in figure \ref{fig1} density (upper) and gas temperature (lower) distributions overlaid with the velocity fields for Model-140 (left panels) and Model-380 (right panels), respectively.
All the physical quantities (i.e., temperature, velocity, etc) except for gas mass density are time-averaged with 
weight of gas mass density during the interval of $t\sim 23800$ -- $24300~{\rm sec}$, while gas mass density is simple time averages with no weight.
Note that Model-140 and Model-380 correspond to the cases with the injection rates of $\dot{M}_{\rm inj} = 350$ and $2000$ ($L_{\rm Edd}/c^2$), respectively (see table 2).

After the simulation starts,
the gas injected from the outer boundary into an initially empty zone
first falls and accumulates around the Keplerian radius,
$r_{\rm K} = 2430 ~r_{\rm S}$,
since the centrifugal force and the gravitational force balance there.
Soon after the transient initial phase accumulated matter spreads outward and inward in the radial direction via viscous diffusion process,
forming an accretion disk extending down to the innermost zone ($t \lesssim 23800 ~{\rm sec}$).
The newly injected matter collides with the disk matter so that a high-density region appears at $\sim(2400$ -- $6000)\ r_{\rm S}$ 
(well outside the Keplerian radius) in figure \ref{fig1}. 
In a sufficiently long time (on the order of the viscous timescale, $t\gtrsim23800~{\rm sec}$, Ohsuga et al. 2005),
quasi-steady, inflow-outflow structure is established (see figure \ref{fig1}).

In figure \ref{fig1} we also indicate the disk surface by the red solid line.
The disk surface was defined in the same way as in K21, with the loci where radiation force balances the gravity in radial direction,
$\chi F_{0,r}/c=\rho GM_{\rm BH}/(r-r_{\rm S})^{2}$.
As in K21, the disks are smoothly connected up to the outer boundary, 
and there is no puffed up structure as seen in the previous RHD simulations\ (see table 1 in K21).

We think that $r_{\rm K} < r_{\rm trap}$ is the only reason to produce a puffed-up structure for the following reason.
When we compare Kitaki et al. (2018) and K21 in which the same code 
was used and $\dot{M}_{\rm BH}$ is not much different, 
only the former (with $r_{\rm K} < r_{\rm trap}$) shows a puffed-up structure, 
while the latter (with $r_{\rm K} > r_{\rm trap}$) not.
The outflow rate in the former is 10 times larger than that of the latter. 
These indicate that the high mass outflow rates obtained in the previous studies 
could be caused by setting a small initial angular momentum (see section 1 of K21).

We understand from the velocity fields in figure \ref{fig1}  that gas is stripped off from the disk surface to form outflow.
We also plot the velocity fields of gas by the white vectors in figure \ref{fig1}.
Near the rotation axis, especially,
we see a cone-shaped funnel filled with high velocity ($\gtrsim 0.3~c$), low density, and high temperature $T_{\rm gas}\gtrsim 10^{8}~{\rm K}$ plasmas
surrounded by the outflow region of modest velocity ($\sim 0.05$ -- $0.1~c$) and modest temperatures,
$T_{\rm gas}\sim10^{6-7}~{\rm K}$.
This velocity and temperature feature is seen in both Model-140 and Model-380.
In both accretion disks, we can see the circular motions (see K21 for detailed analysis).\newline

\subsection{Mass inflow rate and mass outflow rate}
\label{sec-mass-in-out}
\begin{figure*}[t]
  \begin{center}
    \includegraphics[width=160mm,bb=0 0 1479 569]{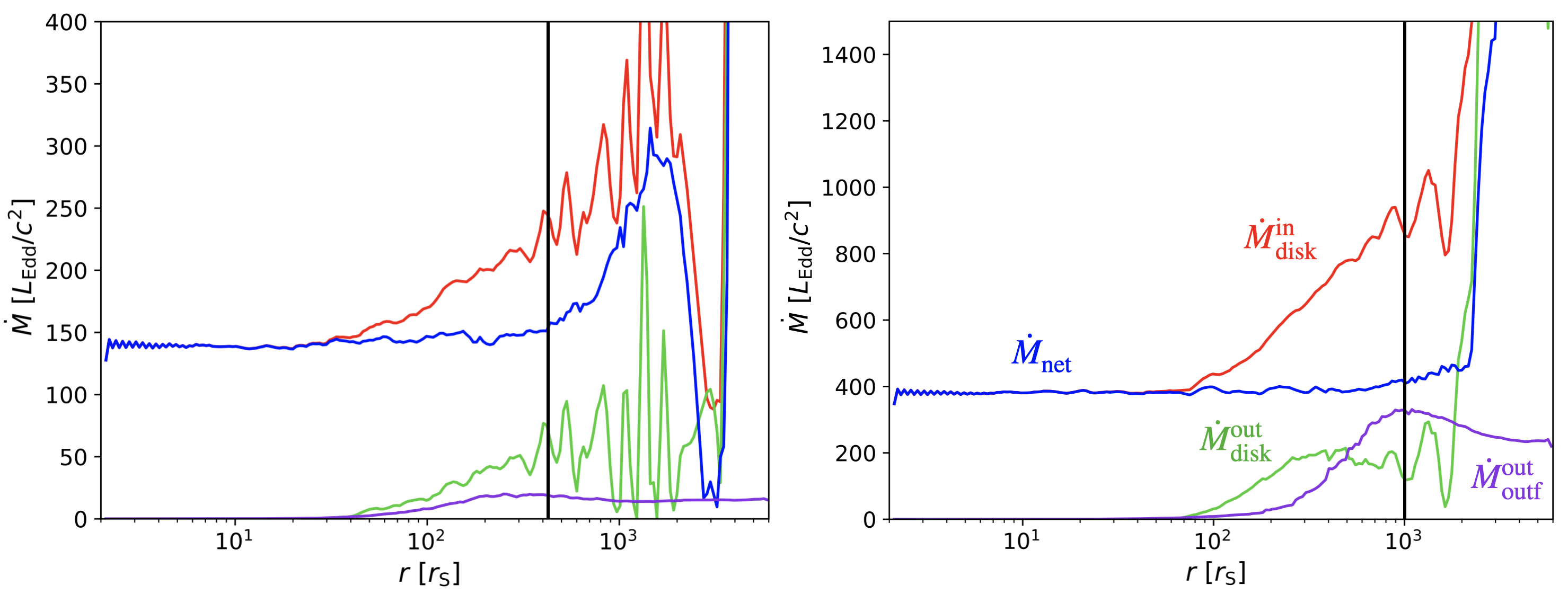}
  \end{center}
  \caption{
    Time-averaged radial profiles of inflow/outflow rate of Model-140 (left panel) and Model-380 (right panel) during the interval of $t\sim 23800$ -- $24300~{\rm sec}$.
    The mass inflow rate within the disk, $\dot{M}_{\rm disk}^{\rm in}$ (red line),
    the mass outflow rate within the disk, $\dot{M}_{\rm disk}^{\rm out}$ (green line),
    the net flow rate, $\dot{M}_{\rm net}$ (blue line),
    and the mass outflow rate in the outflow region (above the disk surface), $\dot{M}_{\rm outf}^{\rm out}$ (purple line), respectively.
    The net flow rate is nearly constant inside the quasi-steady radius, 
    $r_{\rm qss}\sim\rqssf ~r_{\rm S}$ at Model-140 and $\sim\rqsss ~r_{\rm S}$ at Model-380, which is indicated by the vertical black line.
  }
  \label{fig3}
\end{figure*}

\begin{table*}[t]
  \tbl{Radiation and outflow properties}{
  \begin{tabular}{llccccc}
      \hline
      quantity& & Model-110 & Model-130 & Model-140 & Model-180 & Model-380 \\ 
      \hline \hline
      BH ~accretion~rate & $\dot{M}_{\rm BH}$ [$L_{\rm Edd}/c^2$]& $\sim$ 110 & $\sim$ 130& $\sim $140 &$\sim $180& $\sim $ 380\\
      mass injection rate &  $\dot{M}_{\rm inj}$ [$L_{\rm Edd}/c^2$] & 350 & 500 & 700 & 700 & 2000\\
      outflow rate at $r_{\rm out}$ & $\dot{M}_{\rm outflow}$ [$L_{\rm Edd}/c^2$] & $\sim$ 10 & $\sim$ 13 & $\sim$ 16 & $\sim$ 24 & $\sim$ 230 \\ 
      failed outflow rate at $r_{\rm out}$ & $\dot{M}_{\rm failed}$ [$L_{\rm Edd}/c^2$] & 0 & $\sim$ 4 & $\sim$ 8 & $\sim$ 15 & $\sim$ 100 \\ 
      quasi steady-state radius & $r_{\rm qss}$ [$r_{\rm S}$] & $\sim$ 500 & $\sim$ 400 & $\sim$ 400 & $\sim$ 600 & $\sim$ 1000 \\
      pure outflow: inner radius & $R_{\rm pure}^{\rm in} $ [$r_{\rm S}$] & $\sim$ 40 & $\sim $ 40 & $\sim $ 40 & $\sim $ 40  & $\sim$ 40 \\
      pure outflow: outer radius & $R_{\rm pure}^{\rm out}$ [$r_{\rm S}$] & $\sim$ 170  & $\sim 110 $ & $\sim 130 $ &$\sim $ 180 & $\sim $ 480\\ 
      failed outflow: outer radius & $R_{\rm failed}^{\rm out}$ [$r_{\rm S}$] & $\sim$ 170 & $\sim 180 $ & $\sim 210 $ &$\sim 210$& $\sim $ 820 \\
      photon trapping radius & $R_{\rm trap}$  [$r_{\rm S}$] & $\sim$ 300 & $\sim $ 330 &$\sim $ 350& $\sim $ 450 & $\sim 1100 $\\ 
      X-ray luminosity & $L_{\rm X}$ [$L_{\rm Edd}$] & $\sim$ 2.0 & $\sim 2.1$ & $\sim 2.3$ & $\sim $2.4 & $\sim 2.7$\\
      mechanical luminosity & $L_{\rm mech}$ [$L_{\rm Edd}$]  & $\sim$ 0.07 & $\sim 0.09$ & $\sim 0.11$ & $\sim $0.16 & $\sim 0.61$\\
      isotropic X-ray luminosity & $L_{\rm X}^{\rm ISO} (\theta)$ [$L_{\rm Edd}$] & $2.0-4.0$ & $2.3-4.4$ & $2.4-4.8$ & $2.3-4.6$ & $2.1-11$\\
      isotropic mechanical luminosity & $L_{\rm mech}^{\rm ISO}(\theta)$ [$L_{\rm Edd}$] & $0.04-0.18$ & $0.02-0.35$& $0.03 - 0.60$ & $0.04 - 1.4$ & $1.0-8.0$\\
      luminosity ratio & $L_{\rm mech}/L_{\rm X}^{\rm ISO}(\theta)$ &  $0.02-0.04$  & $0.02-0.04$ & $0.02-0.05$ & $0.04-0.07$ & $0.06-0.29$\\ \hline
  \end{tabular}}
  \begin{tabnote}
   Note again that the results of Model-180 are taken from K21.
   In the calculations of isotropic luminosities (see the last three rows), 
   we take an angular range of $0^{\circ} < \theta < \theta_{\rm surf}$.
  \end{tabnote}
  \label{table2}
\end{table*}

It will be of great importance to plot the radial profiles of the mass flow rates
so as to see to what extent a quasi-steady condition is satisfied, and to clarify the gas dynamics around black holes. 
Following K21, we calculate the four flow rates:
the mass inflow and outflow rates within the disk,
the mass outflow rate in the outflow region (the region outside the disk surface),
and the net flow rate:
\begin{eqnarray}
  \dot{M}_{\rm disk}^{\rm in}(r)&\equiv&4\pi\int_{\theta_{\rm surf}}^{\pi/2} d\theta\sin\theta\nonumber\\
  &&~~~~~~~~~\times r^{2}\rho(r,\theta){\rm min}\left\{v_{r}(r,\theta),0\right\},\\
  \dot{M}_{\rm disk}^{\rm out}(r)&\equiv&4\pi\int_{\theta_{\rm surf}}^{\pi/2} d\theta\sin\theta\nonumber\\
  &&~~~~~~~~~\times r^{2}\rho(r,\theta){\rm max}\left\{v_{r}(r,\theta),0\right\},\\
  \dot{M}_{\rm outf}^{\rm out}(r)&\equiv&4\pi\int_{0}^{\theta_{\rm surf}} d\theta\sin\theta\nonumber\\
  &&~~~~~~~~~\times r^{2}\rho(r,\theta){\rm max}\left\{v_{r}(r,\theta),0\right\},   \label{eq-mdotoutflow}\\
  \dot{M}_{\rm net}(r)&\equiv&\dot{M}_{\rm disk}^{\rm in}(r)+\dot{M}_{\rm disk}^{\rm out}(r) + \dot{M}_{\rm outf}^{\rm out}(r).
\end{eqnarray}
Here, $\theta_{\rm surf}=\theta_{\rm surf}(r)$ is the angle between the rotation axis and the disk surface (see the red line in figure \ref{fig1}).

Figure \ref{fig3} illustrates the absolute values of the various mass flow rates as functions of radius, $r$.
We here omit the mass inflow rate in the outflow, since it turns out to be practically zero.

Let us first focus on the case of Model-380\ (see the right panel of figure \ref{fig3}).
The blue line, which represents the net accretion rate, 
provides important information to judge to what extent a quasi-steady state is achieved.
We see that it is approximately constant in the range of $r=(2 - \rqsss)~r_{\rm S}$;
that is, the quasi-steady radius (inside which a quasi-steady state realizes) is $r_{\rm qss}\sim \rqsss ~r_{\rm S}$.

We notice that the mass outflow rate is negligibly small not only in the far outer region but also in the innermost region (see also K21).
We estimate the radius, $R_{\rm pure}^{\rm in }$ ($=R_{\rm inflow}$ in K21),
inside which outflow is negligible, by the intersection of the two lines: 
$\dot{M}_{\rm disk}^{\rm in}(r)$ and $\dot{M}_{\rm net}(r)$,
finding $R_{\rm pure}^{\rm in} \sim 40 ~r_{\rm S}$.
The mass inflow rate in the disk region and the outflow rate in the outer region 
averaged over the range of $r = (2 - 30)~ r_{\rm S}$ are
$\dot{M}_{\rm BH}\equiv\langle|\dot{M}_{\rm disk}^{\rm in}|\rangle=\mdotinaves~L_{\rm Edd}/c^{2}$,
and
$\langle\dot{M}_{\rm outf}^{\rm out}\rangle=\mdotoutaves~L_{\rm Edd}/c^{2}$,
respectively (see also table \ref{table2}).

We are now ready to examine where outflow emerges 
by the examination of the lines in the middle region $(80 - 1000)~r_{\rm S}$.
The outflow rate above the disk surface, $\dot{M}_{\rm outf}^{\rm out}$
(indicated by the purple line in figure \ref{fig3}), increases with increasing radius,
reaches its maximum value of $320~L_{\rm Edd}/c^2$ at 
$r=\rlaus ~r_{\rm S} ~ (\equiv r_{\rm failed}^{\rm out})$, and then decreases beyond.
This position, $r_{\rm failed}^{\rm out}$ ($=r_{\rm lau}$ in K21), corresponds to the outermost launching position of the outflows. 
The fact that $\dot{M}_{\rm outf}^{\rm out}$ decreases beyond $r_{\rm failed}^{\rm out}$ means
that some of the outflow materials fall back onto the disk surface (see K21).
Therefore, this decrement in the $\dot{M}_{\rm outf}^{\rm out}$ curve gives the failed outflow rate.

In the further outer region, $r\gtrsim 3000~r_{\rm s}$, $\dot{M}_{\rm outf}^{\rm out}$ is nearly constant.
The space-averaged (genuine) outflow rate at $r= (4000 - 6000)~r_{\rm S}$ is 
$\dot{M}_{\rm outflow}\equiv\langle\dot{M}_{\rm outf}^{\rm out}\rangle\sim \mdotoutends ~L_{\rm Edd}/c^{2}$.
The total amount of failed outflow is $\dot{M}_{\rm failed}= 100~L_{\rm Edd}/c^2$.

Similar analyses can be repeated for Model-140 (see the left panel of figure \ref{fig3}).
The results of outflow rate evaluations are summarized in table \ref{table2}, 
including those models not plotted in figure \ref{fig3}.

Note that the mass accretion rate onto a black hole (${\dot M}_{\rm BH}$) is determined by mass flow rate at $r=r_{\rm qss}$, and not by the mass injection rate at the outer boundary. 
It thus happens that different accretion rates may appear for the same mass injection rate, as in the case of Model-140 and Model-180.
We should also note that since high-low transitions are observed in Model-110, 
we time-averaged over 200 seconds solely during the super-Eddington state to calculate quantities listed in table 2.


\subsection{Outflow streamlines}
\label{sec-streamline}

\begin{figure*}[t]
  \begin{center}
    \includegraphics[width=160mm,height=100mm,bb=0 0 809 549]{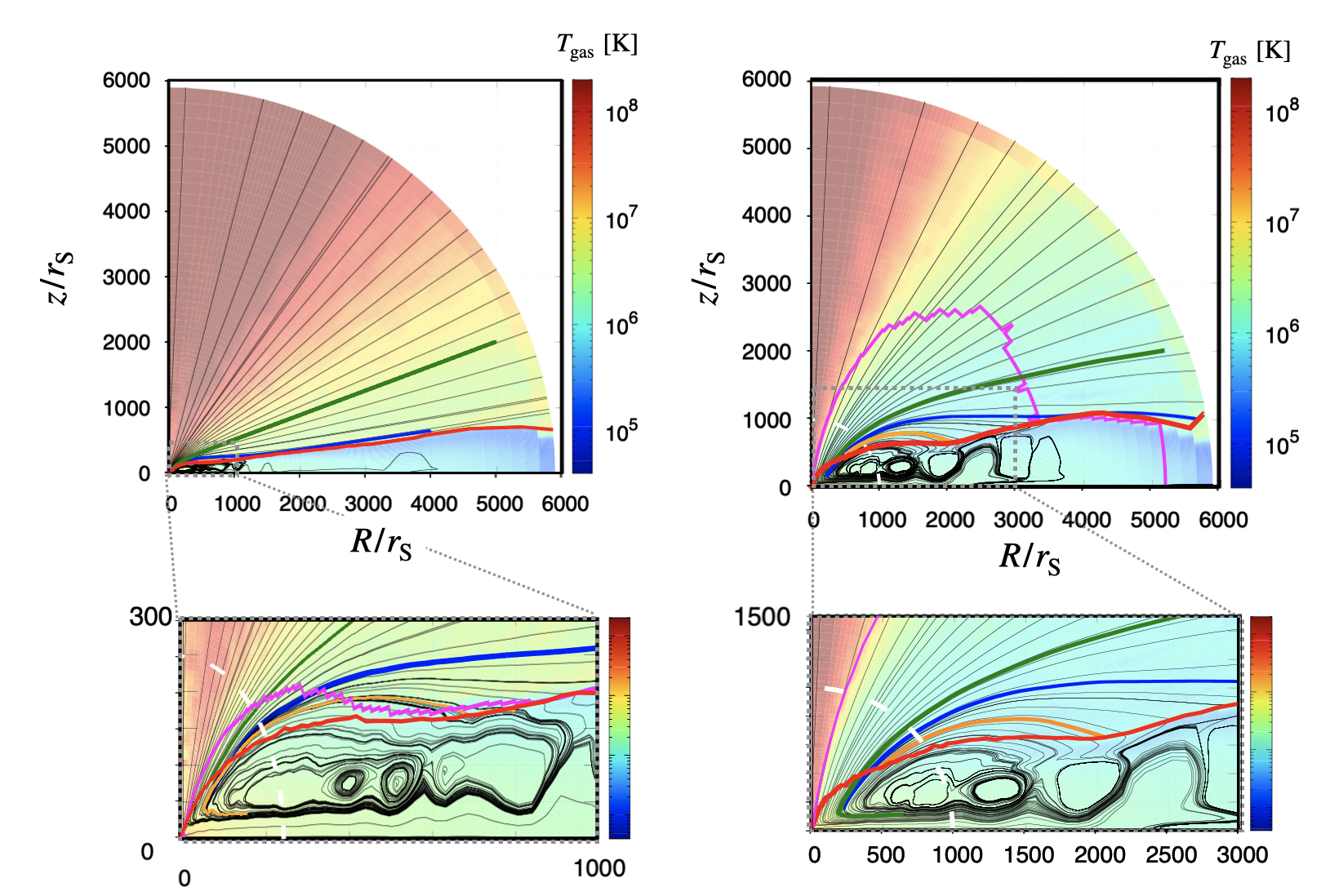}
  \end{center}
  \caption{
    Sequences of streamlines overlaid on the gas temperature distributions for Model-140 (left panel) and Model-380 (right panel), respectively.
    The upper panels are the large-scale view, while the lower ones are the magnification of the central region.
    In each panel we pick up several streamlines and colored them:
    The green line and the orange line indicate a sample streamline in the pure outflow and the same in the failed outflow, respectively, while the blue line shows an interface separating pure and failed outflow regions.  The red line represents the disk surface, and the white line indicates the radius ($r$), at which the cumulative outflow rate reaches its maximum.
}
  \label{fig4}
\end{figure*}

The streamline analysis is useful to understand the outflow path and the evolution of physical quantities of the outflowing gas after being launched.
The upper panel of figure \ref{fig4} displays a sequence of streamlines overlaid on the temperature contours,
while the lower panel is the magnified view of the central region of the upper panel.
We understand in this figure that the outflow emerges from the disk surface 
inside the white circle of the radius, $r \sim \rlaus ~r_{\rm S}$, 
where $\dot{M}_{\rm outf}^{\rm out}$ reaches its maximum (see figure \ref{fig4}).
We, here, define the farthest launching radius $R_{\rm failed}^{\rm out}$ (= $R_{\rm lau}$ in K21)
where the white line crosses the red line; that is $R_{\rm failed}^{\rm out} = r_{\rm failed}^{\rm out}\times\sin{\theta_{\rm surf}}\sim ~1000\times \sin{\left(0.96\right)}~r_{\rm S} \sim\rlaucyls~r_{\rm S}$.

The region between the blue and red lines in figure \ref{fig4} 
indicate the region of the failed outflow;
that is, the outflow which once leaves the disk surface at small radii but eventually comes back to the disk at large radii (see also K21).
The farthest launching radius of the pure outflow (which can reach the outer boundary of the computational box) is given by
 $R_{\rm pure}^{\rm out}=r_{\rm pure}^{\rm out}\times \sin{\theta_{\rm surf}}\sim~650\times \sin{\left(0.72\right)}~r_{\rm S} \sim \rlauinfcyls~r_{\rm S}$ (= $R_{\rm lau}^{\infty}$ in K21).
Here, we define the inner edge of failed outflow, $R_{\rm failed}^{\rm in}$.
Note $R_{\rm pure}^{\rm out} = R_{\rm failed}^{\rm in}$ by definition and this radius is the inner intersection of the blue and red lines in figure \ref{fig4}.

In fact, we see in figure \ref{fig4} that 
the outflow launching from the disk surface at $R_{\rm pure}^{\rm out}\leq R \leq R_{\rm failed}^{\rm out}$ 
does return to the disk surface at larger radii.

As seen in figure \ref{fig4}, outflow launching region can rigorously be identified through the streamline analysis.
As a result, the disk surface can be divided into several regions: 
(1) the innermost region, where outflow rate is negligible, 
(2) the inner region, where pure outflow emerges, 
(3) the middle region, where failed outflow emerges,
and (4) the outer region, where again outflow rate is negligible.
Note that region (3) disappears at very low accretion rate (i.e., Model-110).
These results are consistent with K21.

\begin{figure}[]
  \begin{center}
    \includegraphics[width=85mm,bb=0 0 1027 740]{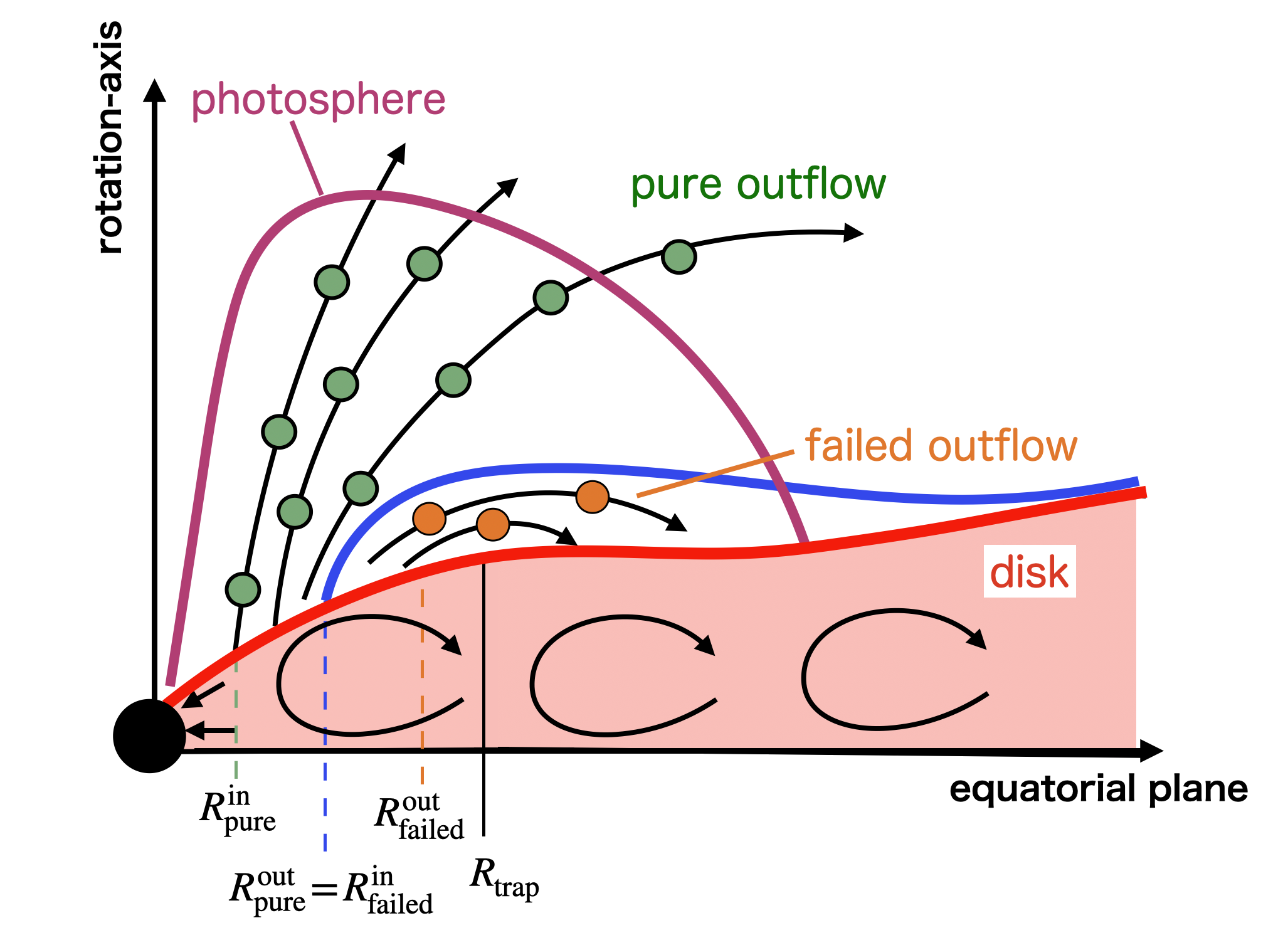}
  \end{center}
  \caption{
  A schematic view of the outflow structure of super-Eddington accretion flow.
  (This figure is basically the same as that shown in K21 except for minor modifications.)
  The pink and blue line represent the photosphere and the interface between the pure outflow region and failed outflow region, respectively.
  The outflow structure is divided into four regions:
  (1) the innermost region with negligible outflow ($R\leq R_{\rm pure}^{\rm in}$),
  (2) the inner region producing pure outflow ($R_{\rm pure}^{\rm in} \leq R \leq R_{\rm pure}^{\rm out}$),
  (3) the middle region producing failed outflow ($R_{\rm pure}^{\rm out} \leq R \leq R_{\rm failed}^{\rm out}$), 
  and (4) the outer region with negligible outflow,
  respectively (see K21).
  }
  \label{kitaki}
\end{figure}

\begin{figure}[]
  \begin{center}
    \includegraphics[width=85mm,bb=0 0 619 461]{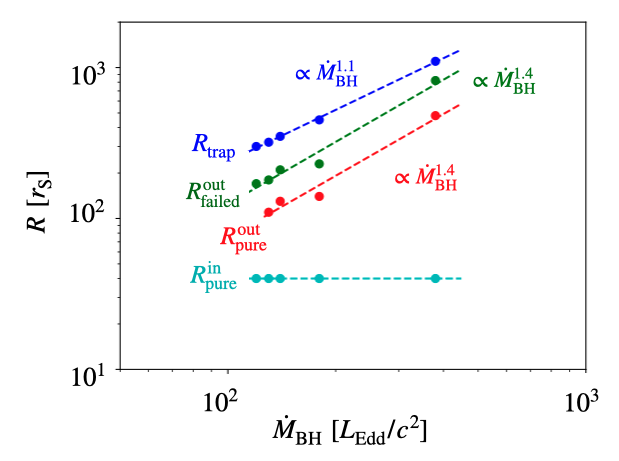}
  \end{center}
  \caption{
    Launching sites of pure and failed outflows as functions of $\dot{M}_{\rm BH}$.
    The cyan, red, green, and blue lines, respectively, represent
    $R_{\rm pure}^{\rm in}$,\ $R_{\rm pure}^{\rm out}$, $R_{\rm failed}^{\rm out}$ and $R_{\rm trap}$, respectively.
    Note that we use the data of K21 to plot in this profile.
  }
  \label{fig5}
\end{figure}

Such outflow structure can be schematically illustrated in figure \ref{kitaki}.
We repeat the same analyses for other models and summarize the results in table \ref{table2}
and in figure \ref{fig5}.
From the fitting we obtain the following scaling laws:
\begin{eqnarray}
  R_{\rm pure}^{\rm out}& \sim &1.9\times10^2\ 
\left(\frac{\dot{M}_{\rm BH}}{200\ L_{\rm Edd}/c^2}\right)^{1.4}~r_{\rm S} ,\\
  R_{\rm failed}^{\rm out} & \sim &3.1\times10^2\
\left(\frac{\dot{M}_{\rm BH}}{200\ L_{\rm Edd}/c^2}\right)^{1.4}~r_{\rm S} ,
\end{eqnarray}
and the photon trapping radius, $R_{\rm trap}$, represents 
\begin{eqnarray}
  R_{\rm trap} &\sim& 5.1\times10^2\ 
\left(\frac{\dot{M}_{\rm BH}}{200\ L_{\rm Edd}/c^2}\right)^{1.1}~r_{\rm S}.
\end{eqnarray}

We notice that the inner boundaries of the launching regions of pure and failed outflows (i.e., $R_{\rm pure}^{\rm out}$ and $R_{\rm failed}^{\rm out}$) 
more steeply with increase in $\dot{M}_{\rm BH}$ than the trapping radius.
This result does not precisely agree with the estimation by the (semi-)analytical model 
(e.g. Shakura \& Sunyaev 1973, Fukue 2004) who show
$R_{\rm pure}^{\rm out} \sim \dot{M}_{\rm BH}/(L_{\rm Edd}/c^2) ~ r_{\rm S}$
(although that they did not distinguish pure and failed outflows; see also K21).

Numerically, our estimations agree reasonably well with theirs at higher accretion rates, e.g.,
${\dot M}_{\rm BH} \sim ~10^2 L_{\rm Edd}/c^2$, 
while ours are much less at lower accretion rates, 
as is shown by the steeper power-law dependence ($\propto {\dot M_{\rm BH}}^{1.4}$) in figure \ref{fig5}.
In other words, outflow emergence region 
(between $R_{\rm pure}^{\rm in}$ and $R_{\rm failed}^{\rm out}$) 
shrinks with a decrease of the accretion rate more rapidly than the simple estimations.

\subsection{Radiation and mechanical luminosities}
\label{Mdot_vs_Luminosity}

\begin{figure}[]
  \begin{center}
    \includegraphics[width=85mm,bb= 0 0 584 431]{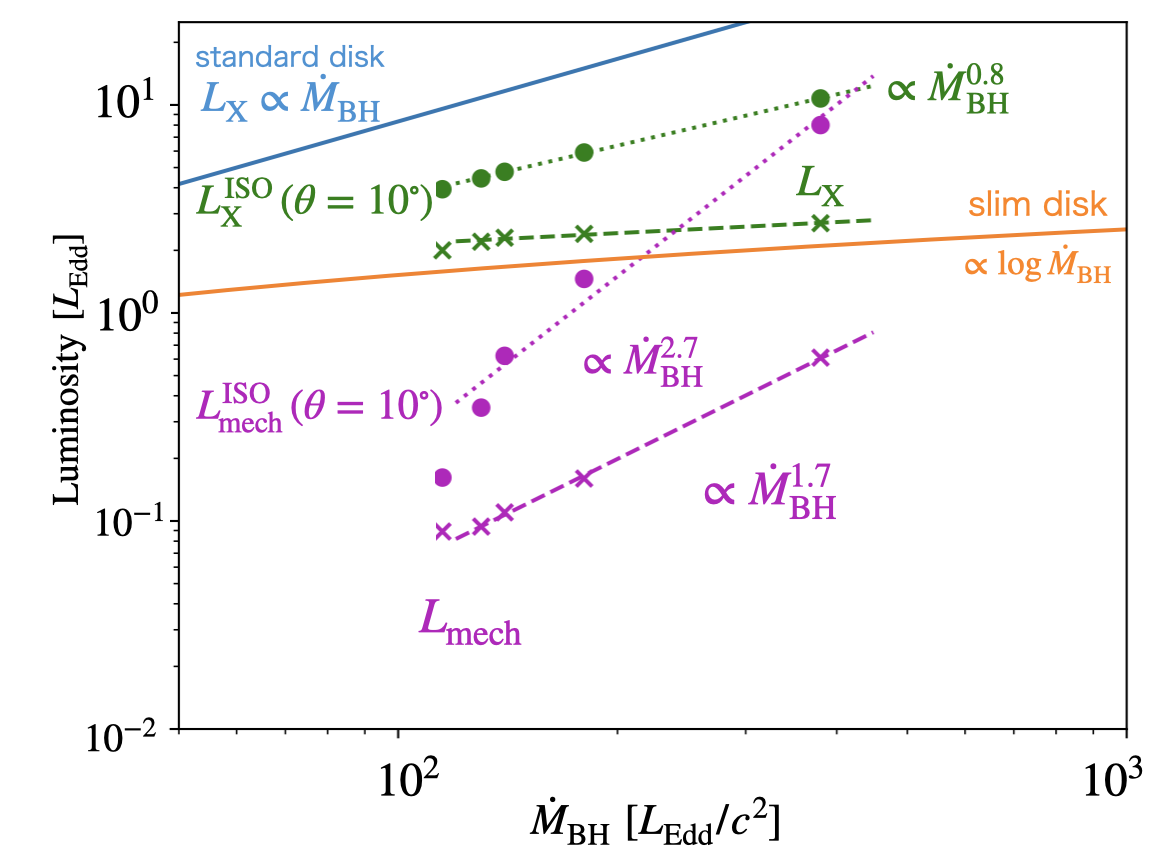}
  \end{center}
  \caption{
    X-ray and mechanical luminosities as functions of $\dot{M}_{\rm BH}$.
    The dashed line and dotted line represent luminosity and isotropic luminosity, respectively.
    For calculation for isotropic luminosity, viewing angle of $\theta = 10^{\circ}$ is assumed.
    The green, magenta, blue and yellow lines represent 
    the radiation and mechanical luminosities calculated by our analyses,
    the radiation luminosity predicted by the standard disk model,
    and the same predicted by the slim disk model, respectively.
  }
  \label{fig6}
  \end{figure}

In this section, we examine how the luminosities depend on mass accretion rate.
We calculate the radiation and mechanical luminosities measured at $r=r_{\rm out}$
by
\begin{eqnarray}
L_{\rm X} &=&4\pi \int_{0}^{\theta_{\rm surf}} \ 
r^2 {\rm max}\{F^{r}_{\rm lab},\ 0\} \sin{\theta} d\theta,\\
L_{\rm mech} &=&4\pi\int_{0}^{\theta_{\rm surf}}\
 r^2 {\rm max}\{\frac{1}{2}\rho v^2 v_r,\ 0 \}\sin{\theta} d\theta.
\end{eqnarray}
Here, $v^2 = v_{\rm r}^2 + v_{\rm \theta}^2 + v_{\rm \phi}^2$
and $F_{\rm lab}^{r}$ is the radial component of radiation flux in the laboratory frame\ (see K21).
$F_{\rm lab}^{r}$ is written as, 
\begin{eqnarray}
  F_{\rm lab}^{i} &=& F_{0}^{i} + v^i E_{0} + v_j P_{0}^{ij}.
  \label{Flab}
\end{eqnarray}
Similarly, the isotropic radiation and mechanical luminosities, defined as 
\begin{eqnarray}
L_{\rm X}^{\rm ISO}(\theta)&=&4\pi r^2 \ {\rm max} \{F^{r}_{\rm lab},\ 0\} ,\\
L_{\rm mech}^{\rm ISO}(\theta) &=&4\pi r^2\  {\rm max}\{\frac{1}{2}\rho v^2 v_r,\ 0\}.
\end{eqnarray}
with $r=r_{\rm out}$, are calculated.
In figure \ref{fig6} we plot $\dot{M}_{\rm BH}$-dependences of different kinds of luminosities.
(We assume the viewing angle of $\theta = 10^\circ$ in this figure.)
We add the blue and orange solid line, which represent the radiation luminosity predicted by the standard disk model 
(Shakura and Sunyaev 1973), and that by the slim disk model (Watarai 2006), respectively.

There are several noteworthy features found in this figure.
First, we focus on the radiation luminosity (the green dashed and dotted lines).
The radiation luminosity (dashed line) shows the $\dot{M}_{\rm BH}$ dependence similar to the slim disk model (the orange solid line).
By contract, the isotropic radiation luminosity (dashed line) depends on $\dot{M}_{\rm BH}$ more sensitively than the (total) luminosity.
For nearly a face-on observer (with $\theta \sim 10^\circ$)
we estimate $L_{\rm X}^{\rm ISO} (10^\circ) \propto \dot{M}_{\rm BH}^{0.8}$. 
Why do the isotropic radiation luminosity and radiation luminosity vary differently? 
This is because of the fact that the radiation energy release is not isotropic, 
and that the higher the accretion rate is, the more focused becomes the radiation flux towards the rotation axis.

Next, we consider the behavior of mechanical luminosity (magenta dashed line and dotted line) with respect to $\dot{M}_{\rm BH}$.
Figure \ref{fig6} shows that mechanical luminosity is more sensitive to $\dot{M}_{\rm BH}$ than X-ray luminosity.
In particular, isotropic mechanical luminosity is found to follow the power-law relation, as $\propto \dot{M}_{\rm BH}^{2.7}$.
We wish to note that the fittings are performed with the exclusion of Model-110.
This is because isotropic mechanical luminosity drops sharply there, as the luminosity approaches the Eddington luminosity, 
at which the radiation force is equal to gravitational force, 
thereby the launch of radiation-pressure driven outflow being suppressed.

Finally, we represent the fitting formula for various luminosities that are valid in the super-Eddington regime, as follows 
\begin{eqnarray}
L_{\rm X} &=& 2.4 \times \left( \frac{\dot{M}_{\rm BH}}{200~L_{\rm Edd}/c^2} \right)^{0.22} ~ L_{\rm Edd},\\ 
L_{\rm mech} &=& 0.20 \times \left( \frac{\dot{M}_{\rm BH}}{200~L_{\rm Edd}/c^2} \right)^{1.7} ~ L_{\rm Edd}, \\
L_{\rm X}^{\rm ISO} (10^{\circ}) &=& 6.4 \times \left( \frac{\dot{M}_{\rm BH}}{200~L_{\rm Edd}/c^2} \right)^{0.83} ~ L_{\rm Edd},\\
L_{\rm mech}^{\rm ISO} (10^{\circ}) &=& 1.4 \times \left( \frac{\dot{M}_{\rm BH}}{200~L_{\rm Edd}/c^2} \right)^{2.7} ~ L_{\rm Edd}.
\end{eqnarray}

We emphasize again the steeper $\dot{M}_{\rm BH}$ dependence of mechanical luminosities than radiation luminosities.
The difference between them is more enhanced, when we consider isotropic luminosities.
For a nearly face-on observer (with $\theta \sim 10^\circ$), especially, 
$L_{\rm mech}^{\rm ISO}(10^\circ)$ becomes comparable to $L_{\rm X}^{\rm ISO}(10^\circ)$ 
at $\dot{M}_{\rm BH}\sim 400~L_{\rm Edd}/c^2$ (or at luminosities of $\sim 10 ~L_{\rm Edd}$).
Such enhanced impact by massive outflow may explain the existence of the anisotropic (elongated) shape of the ULX bubble.
However, we should keep in mind that (unlike the isotropic radiation luminosities)
the isotropic mechanical luminosities are not easy to measure observationally, 
since the impact of the outflow tends to be more or less circularized within a bubble.
We had better to discuss in terms of isotropic radiation luminosities and total mechanical luminosities (see also discussion in section 4.3).

\subsection{Why is isotropic mechanical luminosity so sensitive to accretion rate?}

\begin{figure*}[]
  \begin{center}
    \includegraphics[width=160mm,bb=0 0 1506 545]{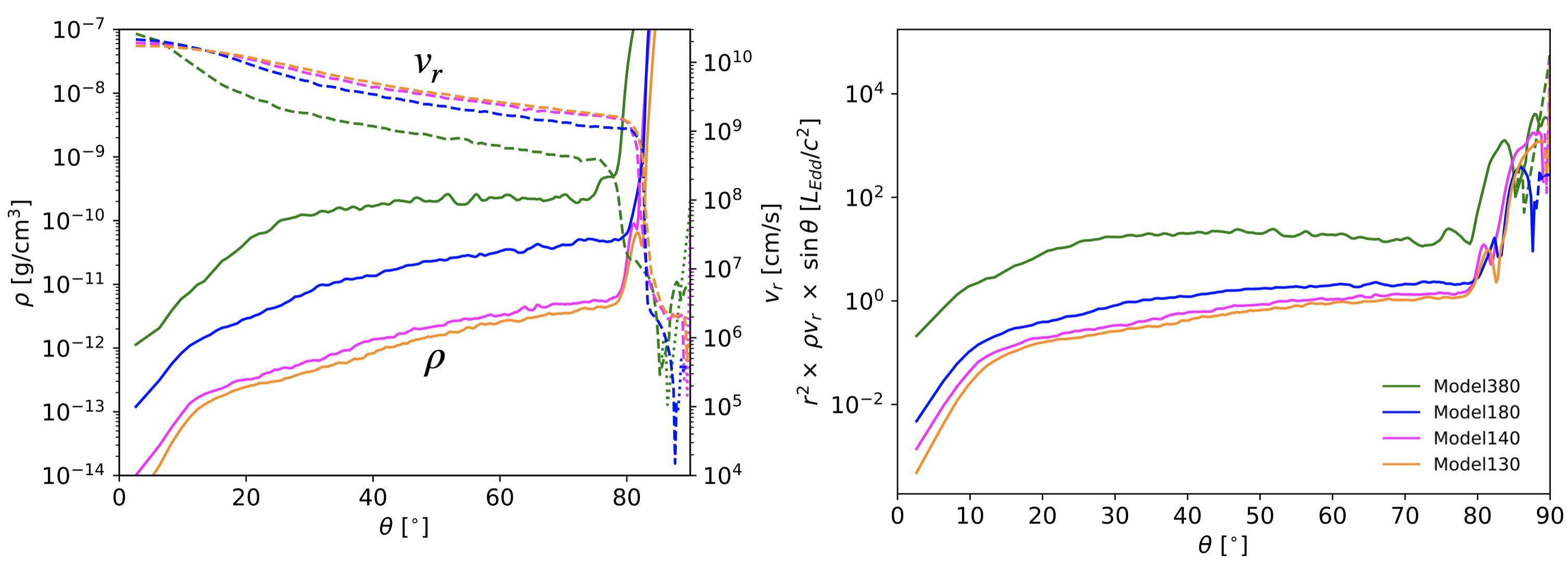}
  \end{center}
  \caption{
   [Left panel] The angular ($\theta$) distributions of the density (solid line) and radial velocity (dashed/dotted line) at $r=5000~r_{\rm S}$for various models: Model-380 (green), Model-180 (blue), Model-140 (pink), and Model-130 (orange), respectively. 
   The dashted and dotted line is outflow and inflow, respectively.
   [Right panel] Same as the left panel but for the distribution of the mass flux (multiplied by $r^2\times \sin(\theta)$).
   The solid and dashed line is outflow and inflow, respectively.
  }
  \label{fig11}
  \end{figure*}

Why does $L_{\rm mech}^{\rm ISO}$ exhibit an extremely large $\dot{M}_{\rm BH}$ dependence? 
Since mechanical luminosity depends on density and radial velocity,
such a rapid growth should be a large increase of either of $\rho$ or $v_r$, or both.
To explore the reason, we plot in the left panel of figure \ref{fig11} the angular ($\theta$) distributions of density and radial velocity at $r=5000~r_{\rm S}$ for various models.
The result is that density increases with increasing $\theta$ for all models, while radial velocity rather decreases.
(As for the angular profiles of the gas density at other radii, see Appendix B.)
In the polar direction, radial velocity does not differ between the models, but density differs significantly.
This gives a direct evidence that the rapid growth of the mechanical luminosity is due to
the rapid increase in density, not in velocity.

We also plot the mass flux multiplied by $r^2 \times \sin\theta$ (right panel of figure \ref{fig11}) at $r=5000~r_{\rm S}$.
This shows that more mass flux goes more preferentially into the intermediate direction than in the polar direction.
We also see that the higher accretion rate is, the more becomes the mass flux profile
(except at very small $\theta$ values).

\begin{figure*}[]
  \begin{center}
    \includegraphics[width=160mm,bb=0 0 971 368]{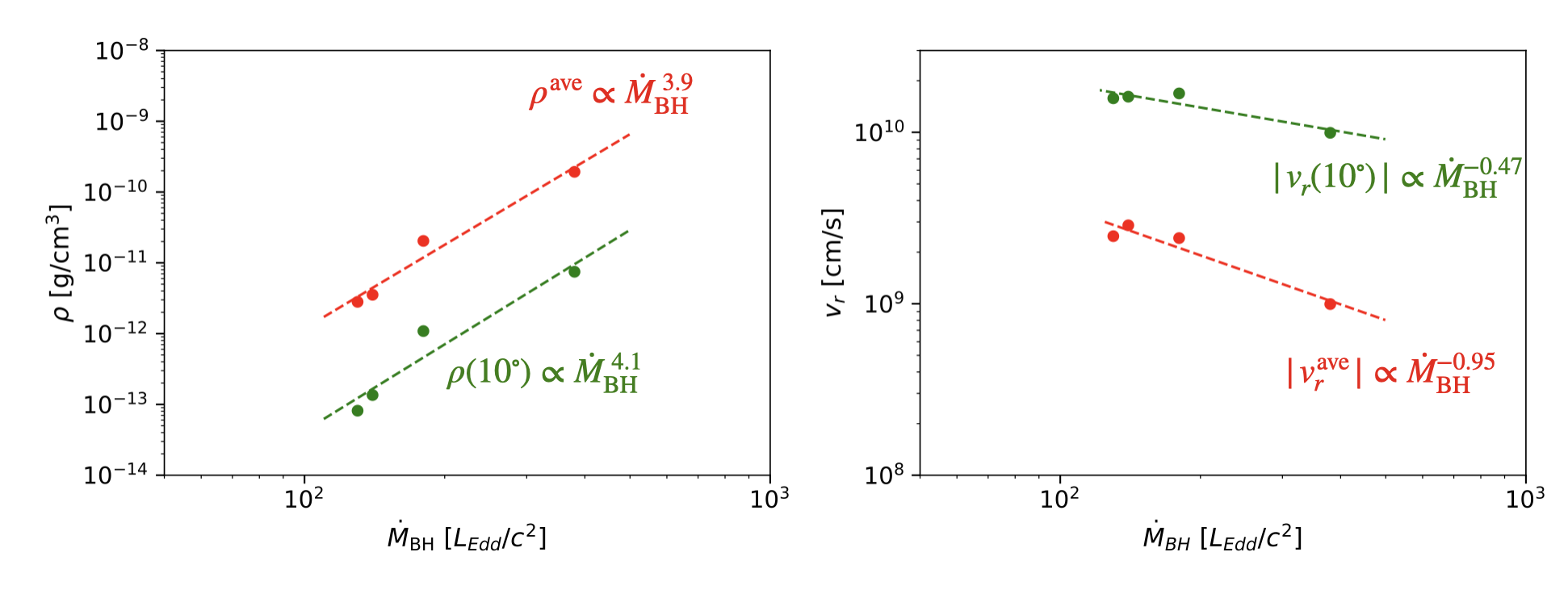}
  \end{center}
  \caption{
    The $\dot{M}_{\rm BH}$ dependence of $\rho$ (left) and $v_r$ (right) at $5000~r_{\rm S}$.
    The red and green lines represent the average value in the flat profile ($\theta=30^\circ-60^\circ$ in Model-380) in figure \ref{fig11}, right panel, and the value in the polar direction ($\theta = 10^\circ$), respectively.
  }
  \label{fig12}
  \end{figure*}

In order to more explicitly demonstrate the rapid growth of density with accretion rates,
we plot the $\dot{M}_{\rm BH}$ dependence of $\rho$ (left) and $v_r$ (right) at $r=5000~r_{\rm S}$ in figure \ref{fig12}.
We define the average values, $\rho^{\rm ave}$ and $v_r^{\rm ave}$ ;
\begin{eqnarray}
  \rho^{\rm ave}& \equiv & \frac{\int_{\theta_{\rm ave}}d\theta \sin{\theta} \times \rho }{\int_{\theta_{\rm ave}} d\theta \sin{\theta}} ,\\
  v_r^{\rm ave}& \equiv & \frac{\int_{\theta_{\rm ave}}d\theta \sin{\theta} \times v_r }{\int_{\theta_{\rm ave}} d\theta \sin{\theta}} .
\end{eqnarray}
Here, $\theta_{\rm ave}$ is angle which mass flux is constant at $5000~r_{\rm S}$ in all models.
The red and green lines represent the average values in the profile in figure \ref{fig11} (left panel), 
and the values in the direction of ($\theta = 10^\circ$), respectively.
We find the following scaling laws:
  \begin{eqnarray}
    \rho (\theta=10^\circ)& \propto &\left(\frac{\dot{M}_{\rm BH}}{200\ L_{\rm Edd}/c^2}\right)^{4.1} ,\\
    v_{r} (\theta=10^\circ)& \propto & \left(\frac{\dot{M}_{\rm BH}}{200\ L_{\rm Edd}/c^2}\right)^{-0.47} .
  \end{eqnarray}

If we use the relationship $L_{\rm mech}^{\rm ISO}\propto \rho \times v_r^3$, 
and if we insert the values at $\theta = 10^\circ$, 
we estimate $L_{\rm mech}^{\rm ISO} \propto \dot{M}_{\rm BH}^{2.7}$,
in reasonable agreement with the result in figure \ref{fig6}.
If we instead adopt the averaged values, we obtain
 $\rho^{\rm ave} \times \left(v_r^{\rm ave} \right)^{3} \propto \dot{M}_{\rm BH}^{~1.1}$,
which does not agree so much.

  \begin{figure*}[]
    \begin{center}
      \includegraphics[width=160mm,bb= 0 0 859 420]{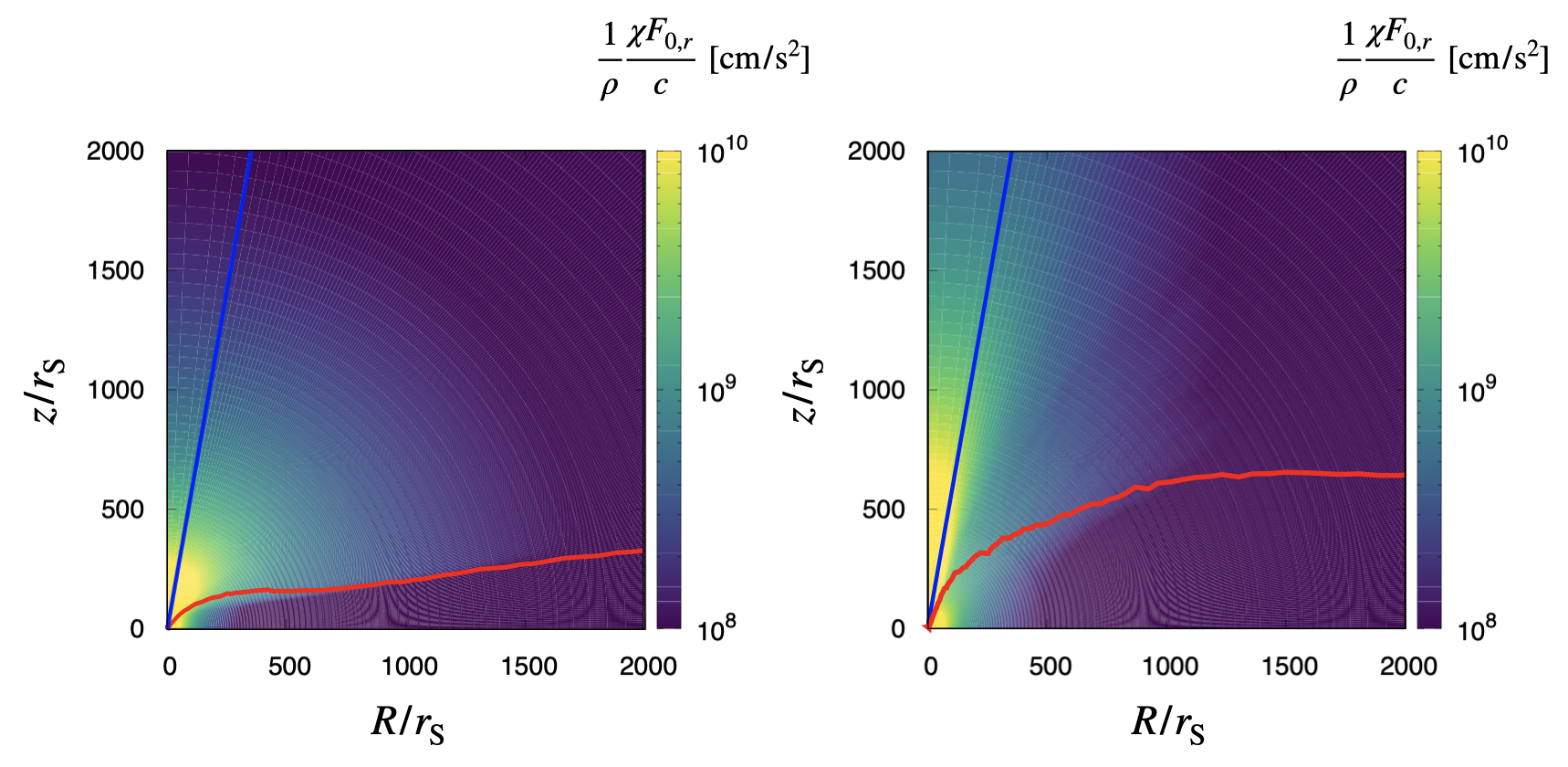}
    \end{center}
    \caption{
    Two dimensional distribution of the radiation force per unit mass 
    (i.e., $\frac{1}{\rho}\frac{\chi F_{0,r}}{c}$) 
    for Model-140 (left panel) and Model-380 (right panel), respectively.
    The red and blue line represent the disk surface and the straight line with $\theta = 10^{\circ}$, respectively.
    We see more collimated high-acceleration region (indicated by the yellow color) in the right panel. 
    This seems to be created due to the self-obscuration, since we see in the right panel
    more vertically inflated disk surface (see the red line standing for disk surface).
    }
    \label{fig13}
    \end{figure*}

Why does the density more rapidly increase towards the polar direction, when mass accretion rates are high?  
To elucidate the reason for this, we plot the distributions of the radial component of the 
radiation force per unit mass, $\chi F_{0,r}/c\rho$, for Model-140 and Model-380 in the left and right panels, respectively, of figure \ref{fig13}.
We there find that the region of strong radiation force per unit mass is more concentrated towards the polar direction, when the mass accretion rate is high (see the right panel).
This seems to be caused by the vertically inflated disk surface, 
which makes radiation field more confined in the region around the rotation-axis, thereby more strongly accelerating outflowing gas.

\section{Discussion}
\subsection{Impact on the environments}
\begin{figure}[]
  \begin{center}
    \includegraphics[width=80mm,bb= 0 0 562 359]{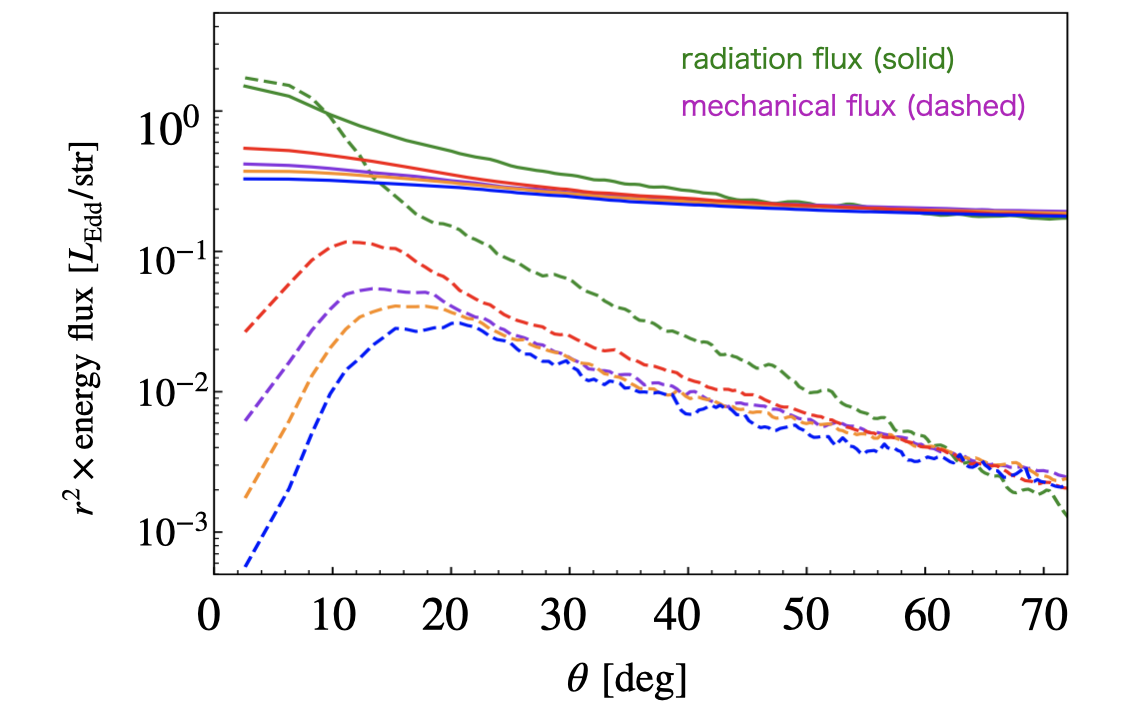}
  \end{center}
  \caption{
   The polar angle ($\theta$) dependences of the radiation (solid) and mechanical (dashed) energy fluxes 
   measured at $r=5000~r_{\rm S}$ for Model-380 (green), Model-180 (red), Model-140 (purple), 
   Model-130 (orange), and Model-110 (blue), respectively.
   Note that the disk surface is located at $\theta = 72^\circ$ in the case of Model-380.
  }
  \label{fig7}
  \end{figure}

It has been suggested that the super-Eddington accretion flow will give large impacts on its environments through powerful outflows.
It is thus crucial to quantify the magnitudes of the impacts from the super-Eddington accretors to properly understand the AGN feedback effects (Botella et al. 2022, King et al. 2003).

Figure \ref{fig7} shows the polar angle ($\theta$) profile of the energy fluxes 
in the laboratory frame (multiplied by $r^2$) measured at $r = 5000 ~r_{\rm S}$ for Model-140 (purple) and Model-380 (green), respectively.
The solid (or the dashed) lines represent the radiation (mechanical) energy fluxes.

Let us first discuss the properties of the radiation energy flux.
We see that the radiation energy flux shows more or less flat profile, but we notice some distinction at small $\theta$ values.
That is, the radiation energy flux steadily grows towards the rotation axis ($\theta = 0$) in Model-380,
whereas it is flatter in Model-140.

We numerically checked the $\theta$ dependence of each term in equation (\ref{Flab}), 
finding that the rapid increase in the energy flux towards the rotation axis, 
which occurs only when accretion rates are large,
is due to the increase of the second and third terms in equation (\ref{Flab}).
We may thus conclude that the distinct shapes of the lines of figure \ref{fig7}
are due to the enhanced advection of the radiation energy within high $\dot{M}_{\rm BH}$ outflow propagating towards the face-on direction.

By contrast, the mechanical energy flux displayed in figure \ref{fig7} exhibit somewhat different behavior; 
all values tend to rapidly grow toward the rotation axis
except in the region close to the rotation axis, where the value turns to decrease in excluding Model-380.
As seen in figure \ref{fig11}, the density curves show similar angular dependence in all models.
The radial velocity profiles, in contrast, exhibit distinct behavior; that is,
the radial velocity in Model-380 rapidly increases toward the polar direction, 
whereas that in the other models only gradually increases.
Because of such somewhat different velocity profile with the different mass accretion rates, 
the impact of the mechanical energy flux on the surroundings becomes more anisotropic, 
as the accretion rate increases.
To summarize, the angular dependence of the energy flux exhibits distinct trends, depending on the accretion rate.

\subsection{The energy conversion}
The energy conversion efficiency is one of the most important key quantities when we discuss the feedback to the environments.
\begin{table*}[]
  \tbl{energy conversion}{
  \begin{tabular}{cccccc}
      \hline
      model & $\dot{M}_{\rm BH} ~ [L_{\rm Edd}/c^2] $& $\dot{M}_{\rm outflow}~[L_{\rm Edd}/c^2]$ & $\beta$ & $\beta_{\rm in}$ & $\beta_{\rm out}$ \\
      \hline \hline
      Model-110 & $\sim 110 $ & $\sim 10$ & 0.11 & 0.92 & 0.08 \\
      Model-130 &  $\sim 130 $ & $\sim 13$ & 0.10 & 0.91 & 0.09 \\
      Model-140 &  $\sim \mdotinavef$ & $\sim 15$ & 0.11 & 0.90 & 0.10 \\
      Model-180 & $\sim 180 $ & $\sim 24$ & 0.14 & 0.88 & 0.12 \\
      Model-380 &  $\sim \mdotinaves$ & $\sim 230$ & 0.61 & 0.62 & 0.38 \\ \hline
    \end{tabular}}
    \begin{tabnote}
      Here, $\beta$ is the ratio of outflow to inflow (equation \ref{eq-ene}) , 
      and $\beta_{\rm out}$ is the ratio of outflow to injected gas from surrounding environment (equation \ref{eq-ene2}).
    \end{tabnote}
    \label{energy-conversion}
\end{table*}
Using the inflow and outflow rates shown in section \ref{sec-mass-in-out}, 
we calculated the inflow and outflow conversion efficiency defined in the same way as in K21; 
\begin{eqnarray}
  \beta&\equiv&\frac{\dot{M}_{\rm outflow}}{\dot{M}_{\rm BH}}\label{eq-ene},\\
  \beta_{\rm in}&\equiv&\frac{\dot{M}_{\rm BH}}{\dot{M}_{\rm BH}+\dot{M}_{\rm outflow}},\label{eq-ene1}\\
  \beta_{\rm out}&\equiv&\frac{\dot{M}_{\rm outflow}}{\dot{M}_{\rm BH}+\dot{M}_{\rm outflow}}. \label{eq-ene2}
\end{eqnarray}
Here, the denominators of equations (\ref{eq-ene1}) and (\ref{eq-ene2}),
$\dot{M}_{\rm BH} + \dot{M}_{\rm outflow}$ mean the injected mass flow rate from surrounding environment under the assumption
that the net flow rate is entirely constant in radius.
Since this assumption is not entirely justified, this value is not precisely equal to ${\dot M}_{\rm inj}$ listed in table 1.
So $\beta_{\rm out}$ represent how much fraction of the injected gas turns into outflow.
We find that 38\% of the injected gas can be converted to outflow when $\dot{M}_{\rm BH} \sim 380 ~L_{\rm Edd}/c^2$. 
We summarize the results in table \ref{energy-conversion}, which shows
that mass inflow with higher rates can be more efficiently converted to outflow than otherwise.

In parallel with the present work,
Hu et al. (2022) performed series of large-scale
and long-term simulations of super-Eddington accretion flows,
adopting various boundary conditions under the optically thick limit, 
and obtained a larger $\beta$ value; 
e.g., $\beta\sim 32.9$ for ${\dot M}_{\rm BH}\sim 311~ L_{\rm Edd}/c^2$.
In addition, they showed a gentler ${\dot M}_{\rm BH}$-dependence 
of the momentum flux, ${\dot P}_{\rm mom}$; that is, roughly
${\dot P}_{\rm mom}\propto {\dot M}^{1}$, 
while our results show ${\dot P}_{\rm mom}\propto {\dot M}^{2}$. 
Numerically, their value is about six times larger at ${\dot M}_{\rm BH}\sim 100~ L_{\rm Edd}/c^2$, 
and about twice at ${\dot M}_{\rm BH}\sim 380 ~L_{\rm Edd}/c^2$.
Hu et al. (2022) claimed that the reasons for the differences from K21, in which the same method is employed as the present work, 
are due to (1) the assumption of equatorial plane symmetry
and to (2) the large alpha parameter in our calculations. 
The cause of the difference will be studied in future work.
We here point out that the outflow region is not entirely optically thick
(for absorption) when mass accretion rate is low
so that the equality between the radiation energy and gas energy density may not always hold.

\subsection{Connection with observations of ULXs}
We next discuss observational implications of our model.
A good target is the ULXs, since they are occasionally associated with optical nebula 
and/or radio bubbles (e.g., Kaaret et al. 2004), 
and since these nebulae are thought to originate from the outflow in super-Eddington accretion flow (Hashizume et al. 2015).
The X-ray luminosity via direct observations of the central objects can be evaluated by us,
while the mechanical luminosity can be estimated by observing optical radiation from the ULX bubble.
As K21 have already pointed,
the ratio of $L_{\rm mech}/L_{\rm X}^{\rm ISO}(\theta)$ should be a good indicator to 
discriminate whether the central object of ULX is a black hole or a neutron star.
In table \ref{table2} we summarize the ratios estimated based on our simulations.
In Model-380, for example, the ratio ranges between 0.06 and 0.29, depending on the angle, $\theta$.
We find somewhat smaller values in other models, but at least we may conclude that these values are
consistent with the observations, which is $0.04-0.14$ for Holmberg  ${\rm II}$ X-1 (Abolmasov et al. 2007)

\subsection{Bernoulli parameter along streamlines}

\begin{figure*}[]
  \begin{center}
    \includegraphics[width=160mm,bb=  0 0 767 568]{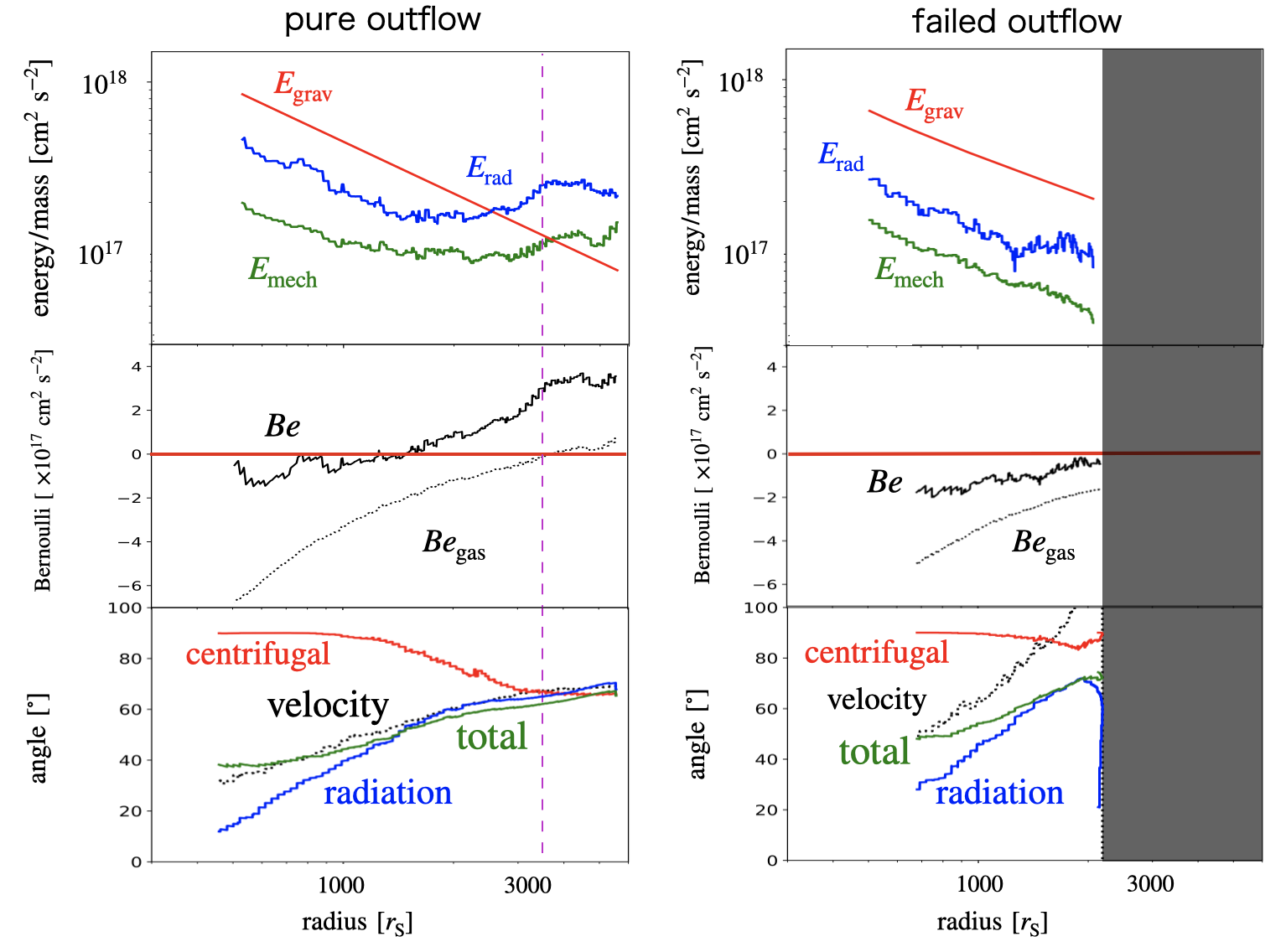}
  \end{center}
  \caption{
    [Top] Energy distribution along one streamline in the pure outflow (left panel) and another in the failed outflow (right panel). 
    The red, blue, and green lines represent the gravitational energy, 
    the radiation energy, and the mechanical energy, respectively.
    The magenta line indicates the position of the photosphere.
    [Middle] Same as the top panel but for the specific Bernoulli parameter of the gas + radiation component (solid line) and of the gas component only (dashed line).
    [Bottom] Same as the top panel but for the polar angles of the velocity vector (black), the total force vector (green), the radiation force vector (blue), and the centrifugal force vector (red), respectively.
  }
  \label{fig8}
  \end{figure*}

In order to investigate what factor is responsible for separating failed outflow and pure outflow, 
we calculate the energy distribution along the respective streamlines of pure and failed outflows.
The top panel of figure \ref{fig8} shows energy distribution along the two representative streamlines 
which are shown in figure \ref{fig4} as the green and orange lines;
the former corresponds to the pure outflow (left panel), while the latter to the failed outflow (right panel).
The red line ($E_{\rm grav}$), blue line ($E_{\rm rad}$), and green line ($E_{\rm mech}$) 
in each panel represent the gravitational energy ($GM/(r-r_{\rm S})$), 
the radiation energy ($E_0/\rho$), and the mechanical energy ($v^2/2$), respectively.
In both flows, the gravitational energy dominates over others at the launching point.
In the left panel, however, the kinetic energy eventually exceeds the gravitational energy during the course of outflow propagation, thereby producing pure outflow.
In the right panel, by contrast, the kinetic energy is entirely less than the gravitational energy so that the failed outflow should appear. 
Only when outflowing gas travels in the region with large $E_{\rm rad}$ for a certain time it can become pure outflow. 

It was previously suggested in the context of optically thin ADAF (advection-dominated accretion flow) that 
the specific Bernoulli parameter could be a good indicator to judge whether or not (pure) outflow can emerge (e.g. Narayan \& Yi 1994). We, here, calculate 
the specific Bernoulli parameter including radiation component,
\begin{eqnarray}
  Be &=& \frac{1}{2}v^2 + \frac{e_{\rm gas}}{\rho} + \frac{E_{0}}{\rho}+ \frac{p}{\rho}+ \frac{P_{\rm rad}}{\rho} - \frac{GM}{r-r_{\rm S}}.
\end{eqnarray}
and that of gas component only,
\begin{eqnarray}
  Be_{\rm gas} &=& \frac{1}{2}v^2 + \frac{e_{\rm gas}}{\rho} + \frac{p}{\rho} - \frac{GM}{r-r_{\rm S}}.
\end{eqnarray}
Here, $P_{\rm rad}$ is the radiation pressure and is
$P_{\rm rad} \equiv \left(\bm{{\rm P}}_{0}^{rr}+\bm{{\rm P}}_{0}^{\theta \theta} +\bm{{\rm P}}_{0}^{\phi\phi}\right)/3$.

The Middle panel of figure \ref{fig8} show the distribution of the specific Bernoulli parameter
of gas component only (dotted line) and the total one (solid line), respectively.
We first notice that they are not constant but increase as outflow propagates, 
since the gas is continuously accelerated by the radiation force.
We also find that the specific Bernoulli parameters are negative at the launching points
in both flows shown here.
A difference is found in that the Bernoulli parameter of gas only can eventually become positive
at around the position of the photosphere in pure outflow, while it never becomes positive in failed outflow.
Thus, we conclude the condition for pure outflow is
that the Bernoulli parameter of gas only can become positive before reaching the photosphere.
To summarize, our simulations demonstrate that pure outflow emerges, even if $Be < 0$ near the launching point.

Finally, we examine the bending of the streamline of pure and failed outflow.
The bottom panel of figure \ref{fig8} show the directions of some representative vector quantities
along the streamline of pure (left panel) and failed (right panel) outflow as functions of $r$:
the velocity (dotted line), the total force (green solid line), 
the radiation force (blue solid line), and the centrifugal force (red solid line).
Here, by the centrifugal force we mean the combination of the second and third term on the right-hand side of equation (\ref{eom_r}) and the second term on the right-hand side of equation (\ref{eom_th}),
and by the total force we mean the sum of the radiation force, centrifugal force, and gravitational force. Note that the gas pressure force is negligible.

We see in this figure that the radiation force is mainly upward (with small angles) at the launching point,
whereas the centrifugal force is in the $R$-direction ($\theta=90^\circ$).
Gas is thus initially accelerated in the intermediate direction ($\theta \sim 40^\circ$)
in both cases of pure and failed outflow.
Within the pure outflow (see the lower left panel) total force, radiation force and velocity vectors tend to direct the same direction
(with $\sim 70^\circ$) and eventually all the angles coincide with each other.
This means that gas dynamics is governed by radiation.
No such converging behavior is observed in the failed outflow (see the lower right panel).
Since gravitational energy does always exceed the radiation energy,
the angle of velocity vector steadily increases and eventually falls down onto the disk surface.

\subsection{The transition in the thermal instability}
\label{transition}
First of all, 
we wish to point that the transitions between the super-Eddington and sub-Eddington states (in our Model-110) and 
the transition to the super-Eddington state shown by Inayoshi et al. (2016) are caused by entirely distinct mechanisms.
Inayoshi et al. (2016) considered a region far away from the black-hole accretion disk (slightly inside the Bondi radius).
In their work, HII gas in the central region, ionized by UV radiation, pushes the outer HI gas by gas pressure gradient forces. 
If the mass density of the interstellar gas is high enough, the gravity exceeds the gas pressure gradient force and the gas cannot be prevented from falling. 
Thus, the HI gas accretes at the supercritical rate.

On the other hand, we focus on the accretion disk much closer to the black hole. 
The cause of the transition appearing in our Model-110 is the thermal instability of the disk. The heating (cooling) rate exceeds the cooling (heating) rate, 
causing a runaway temperature increase (decrease), leading to the significant change in the mass accretion rate. 
According to the disk instability theory (see, e.g., Chap. 10 of Kato et al. 2008), 
a thermal instability occurs outside the trapping radius in the case that the dynamical viscosity is proportional to the total pressure.

In models other than Model-110, no such instabilities are observed, 
probably because the spatial extent of the unstable region is limited between the trapping radius and 
the quasi-steady radius and both radii are closer to each other; that is, $r_{\rm qss}/r_{\rm trap} \sim  0.9 - 1.3$ in other models 
(note $r_{\rm qss}/r_{\rm trap} \sim 1.7$ in Model-110, see table 2).
 As a consequence, a thermal instability, even if it may occur locally, cannot propagate widely to produce global, coherent state transitions. 
 If we could increase the trapping radius, we would be able to obtain state transitions, but such a study is beyond the scope of the present paper and is left as future work.

\section{Concluding remarks}
In the present study, we perform extensive radiation-hydrodynamics simulations for a variety of mass injection (and mass accretion) rates to see how the properties of radiation and outflow depend on the input parameter. 
The specific questions that we have in mind are two-fold:
(Q1) How do the radiation and mechanical luminosities depend on $\dot{M}_{\rm BH}$ and inclination angle, 
and (Q2) how much material is launched from which radii and to which directions?

  In order to avoid numerical artefacts as much as possible and to precisely evaluate the impacts from super-Eddington accretors, we set relatively large calculation box with box size of 6000 $r_{\rm S}$ (or $3000~r_{\rm S}$)
and assume relatively large Keplerian radii (2430 $r_{\rm S}$).
We have the following results, some of which are unexpected before the present study.
\begin{itemize}
\item
  We find that the mechanical luminosity grows more rapidly than 
  the radiation luminosity with an increase of ${\dot M}_{\rm BH}$.
\item
 Since the isotropic mechanical luminosity ($\propto {\dot M}_{\rm BH}^{2.7}$) grows much faster
than the isotropic radiation luminosity ($\propto {\dot M}_{\rm BH}^{0.8}$),
the ratio, $L_{\rm mech}/L_{\rm X}^{\rm ISO}$, steadily increases as accretion rate increases.
They could be comparable (and are $\sim 10~L_{\rm Edd}$) for $\theta=10^{\circ}$ 
at the accretion rate of ${\dot M_{\rm BH}} \sim 400~L_{\rm Edd}/c^2$.
\item
  We examined which factor is essential to produce such a rapid growth with accretion rate,
  finding that it seems to be caused by the vertically inflated disk surface, which makes radiation field more confined in the region around the rotation-axis, 
  thereby more strongly accelerating outflowing gas.
\item
  There are two kinds of outflow: pure outflow and failed outflow.
  We find that the fraction of the failed outflow decreases as the accretion rate decreases,
  and that no obvious failed outflow is observed when ${\dot M}_{\rm BH}= 110 ~L_{\rm Edd}/c^2$.
\item
  The higher ${\dot M}_{\rm BH}$ is, the larger become the ratio of the outflow to the inflow ($\beta$) and the launching radii ($R_{\rm pure}^{\rm in}$ and $R_{\rm failed}^{\rm in}$).
  Roughly, $R_{\rm pure}^{\rm out}\propto{\dot M}_{\rm BH}^{1.4}$.

\item
 The angular profile of the outflow is nearly flat except near the rotation axis,
while the magnitude of the impact (energy and momentum) grows towards the rotation axis.
This is because of rapid growth of $v_r$, which counteracts decrease of $\rho$. 

\item
  We investigate physical quantities along outflow trajectories, 
  finding that the Bernoulli parameter is no longer a good indicator to discriminate pure and failed outflows.
  In fact, pure outflow can arise, even if $Be < 0$ at the launching point.

\item
  The motivation for introducing a small injection angle (mass injection area) is to reduce 
  as much as possible the impact of the inflow on the outflow in the computational domain. 
  More vertically inflated structure could appear, if the angle of injection region is larger than that of the disk. 
  Even in such cases, however, the resultant outflow properties will not alter significantly, 
  since the direction of the outflow mechanical energy flux is not towards the equatorial plane but towards the region of relatively small polar angles. 
  As future work, we wish to consider the effects of changing the mass injection angles in a more quantitative fashion to examine.

\item
    When we decrease the mass injection rate, we expect the oscillations of a sort similar to those of Model-110 to occur. 
    Both of the total and the isotropic radiation luminosities will decrease, as the decrease of mass accretion rate, 
    but their separation will tend to reduce, since the discrepancy is caused by the particular geometrical shape of the disk 
    (which tends to confine the radiation field in the polar direction) only at high luminosity state.
    By contrast, the total and isotropic mechanical luminosities will vanish, as the radiation luminosity approaches the Eddington luminosity.

\item
  As future issues we need to solve the magnetohydrodynamics, since then MHD driven outflow will appear and may partly modified the radiation-driven outflow. General relativistic calculations are another issue to be incorporated.
  We then simulate the Blandford-Znajek type jet (outflow) in addition (Blandford \& Znajek 1977).

    It has been suggested in the 3D GR-RMHD simulations of subcritical accretion flows, 
  in addition, that a puffed-up disk vertically predominantly supported by the magnetic pressure (Lancova et al. 2019). 
  If we would run RMHD simulations we may find a more vertically puffed-up structure than in the present study, but this is also left as a future work.

\item 

    Finally, we mention the GR effects.
    We expect that the main results would not qualitatively change for the case of a Schwarzschild black hole. 
    In the rapidly spinning Ker hole, in contrast, the BZ effect causes energy injection through the Poynting flux into the gas near the black hole, 
    which will lead to significant enhancement of the mechanical power of the outflow (Narayan et al. 2017, 2022, Sadowski et al. 2014, Utsumi et al. 2022).

    Another GR effect is found in the radius of the inner edge of the disk (i.e., ISCO radius).
    Tchekhovskoy \& McKinney (2012) performed the GR-MHD simulations of the flow around rapidly spinning black holes 
    with the spin parameter of $a = -0.9$ and $a = +0.9$, finding more powerful outflow in the latter than the former (see also Utsumi et al. 2022).
\end{itemize}

\begin{ack}
  This work was supported in part by JSPS KAKENHI grant JP18K13594 (T.K.), 
  JSPS Grant-in-Aid for Scientific Research (A) JP21H04488 (K.O.), the same but for Scientific Research (C) JP20K04026 (S.M.), and JP18K03710 (K.O.). 
  This work was also supported by MEXT as “Program for Promoting Researches on the Supercomputer Fugaku” 
  (Toward a unified view of the universe: from large-scale structures to planets, JPMXP1020200109; K.O., and T.K.), 
  and by Joint Institute for Computational Fundamental Science (JICFuS; K.O.).  
  Numerical computations were in part carried out on Cray XC50 at Center for Computational Astrophysics, National Astronomical Observatory of Japan.
\end{ack}

\clearpage
%
\appendix
\setcounter{section}{0} 
\renewcommand{\thesection}{\Alph{section}} 
\setcounter{equation}{0} 
\renewcommand{\theequation}{\Alph{section}\arabic{equation}}
\setcounter{figure}{0} 
\renewcommand{\thefigure}{\Alph{section}.\arabic{figure}}
\setcounter{table}{0} 
\renewcommand{\thetable}{\Alph{section}\arabic{table}}

\section{Light curves}

\begin{figure}[]
  \begin{center}
    \includegraphics[width=80mm, bb= 0 0 439 470]{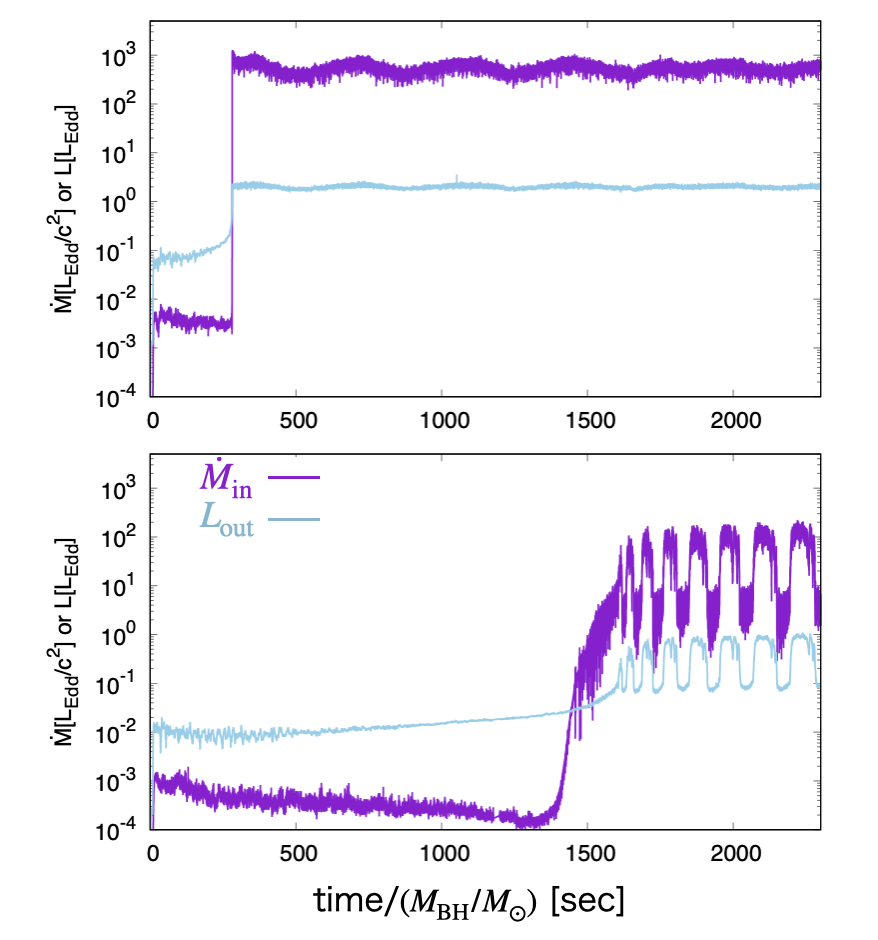}
  \end{center}
  \caption{
    Time evolution of the mass input rates and outgoing radiation luminosities for Model-380 (upper panel) and Model-110 (bottom panel).
    The purple and blue line represent the mass accretion rate $\dot{M}_{\rm in}$ and the luminosity $L_{\rm out}$.
    The luminosities are smaller than the values in table 2, since these are calculated from the first step simulation data,
    in which the inner edge is set to be at $20 ~r_{\rm s}$.
  }
  \label{figa1}
\end{figure}

In the present study, the RHD simulations were performed in 
two steps. In the first step, we set the radius of the inner boundary 
to be at $20~r_{\rm S}$. This is to save computational time. After confirming 
that the flow settles down to a quasi-steady state down to $20~ r_{\rm S}$, 
we start second step simulations by using the data of the first-step 
simulations, but setting the inner boundary to be at $2~r_{\rm S}$. 
All the analyses and figures presented in sections 3 and 4 were made 
by using the second-step simulation data.

In figure \ref{figa1} we show time evolution of $\dot{M}_{\rm  in}$ (purple line) and 
$L_{\rm out}$ (blue line) for Model-380 (upper panel) and Model-110 
(lower panel) in the first-step simulations, respectively. 
We calculate the accretion rate at $r = 20~r_{\rm S}$ and the luminosity 
at $r = r_{\rm out}$ by 
\begin{eqnarray}
  \dot{M}_{\rm in} & \equiv & 4\pi\int_{0}^{\pi/2} d\theta\sin\theta \times \left(20~r_{\rm S}\right)^{2}\nonumber\\
  &&~~~~~~~~~\times \rho(r,\theta)~|{\rm min}\left\{v_{r}(20~r_{\rm S},\theta),0\right\}|,\\
  L_{\rm out}& \equiv & 4\pi\int_{0}^{\pi/2} d\theta\sin\theta \times r_{\rm out}^2 {\rm max}\left\{F_0^r,0\right\}.
\end{eqnarray}
We should note that the definitions and the absolute values of $\dot{M}_{\rm in}$ and $L_{\rm out}$ are slightly 
different from $\dot{M}_{\rm BH}$ and $L_{\rm X}$ used in the text. 
In Model-380, the mass accretion rate abruptly increases from $10^{-3}$ to $10^{2}$ ($L_{\rm Edd}/c^2$), 
when the injected gas reaches the inner radius, and then settles down to a stable super-Eddington state.
These features are consistent with those of the previous studies.

In Model-110, high-low transitions between the super-Eddington state and the sub-Eddington state are observed with a constant interval. 
This is the same sort of the limit cycle oscillations (e.g., Abramowicz et al. 1988, Honma et al. 1991, Ohsuga 2006, 2007).
It occurs due to a thermal instability which occurs when radiation pressure is dominated.
No such transitions are observed in other models (this issue was discussed in section \ref{transition}).

\setcounter{figure}{0} 
\renewcommand{\thefigure}{\Alph{section}.\arabic{figure}}

\section{Angular profiles of density}
\begin{figure}[]
  \begin{center}
    \includegraphics[width=80mm,bb=  0 0 649 472]{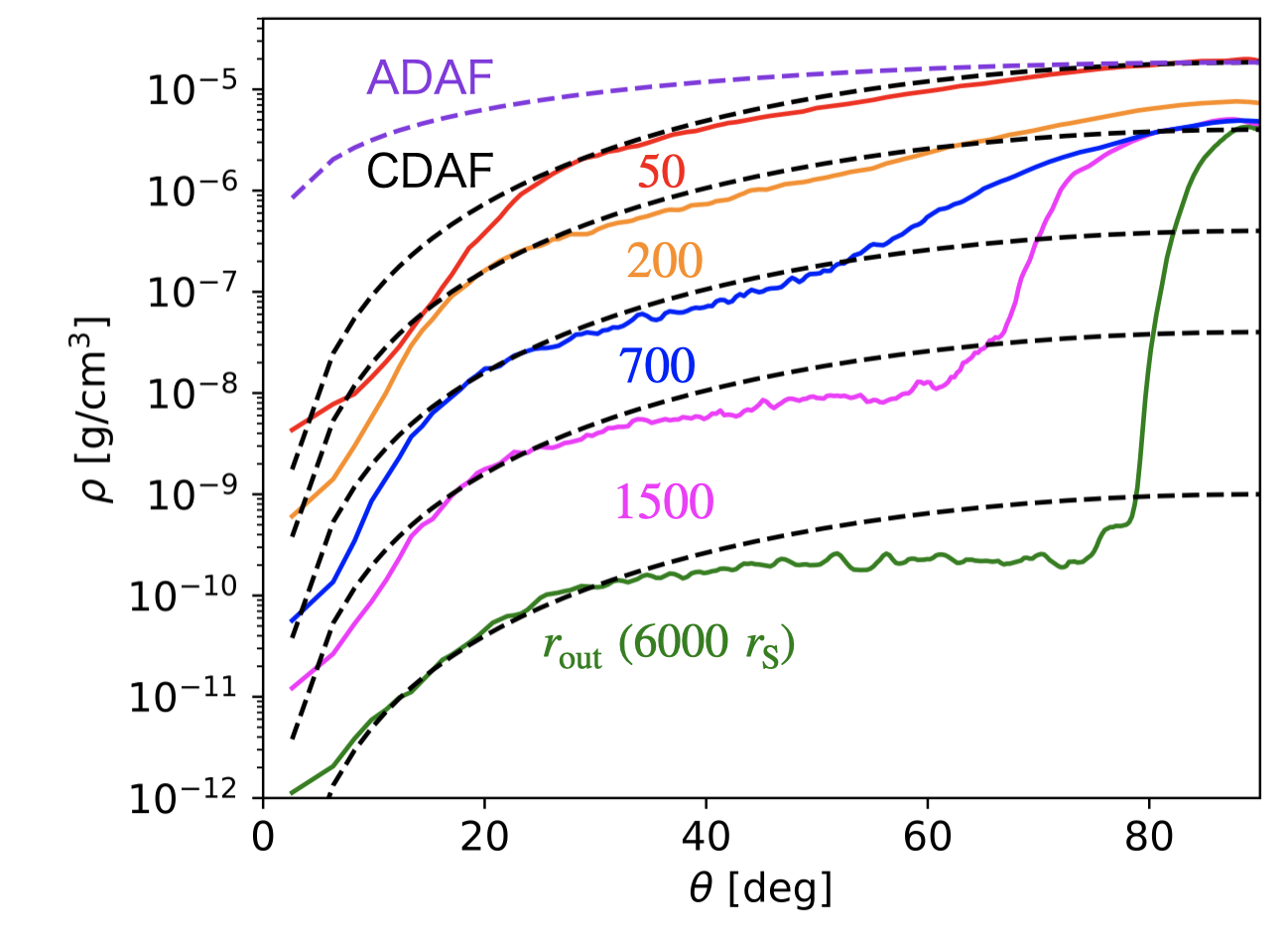}
  \end{center}
  \caption{
    The angular ($\theta$) prifiles of the density.
    The solid lines represent the simulated profiles at various radii: $50~r_{\rm S}$ (red), $200~r_{\rm S}$ (orange), $700~r_{\rm S}$ (blue), 
    $1500~r_{\rm S}$ (pink) and $6000~r_{\rm S}$ (green), respectively.
    The dashed lines represent those of CDAF (black) and ADAF (purple), respectively.
    }
  \label{figb1}
  \end{figure}
  It may be interesting to compare our results with those of the CDAF or ADAF. We plot the angular density profiles at several radii 
  in figure \ref{figb1} (by the solid lines), together with those of the CDAF and ADAF (Quataert et al. 2000).  
  The black dashed line represents the CDAF solution (with $n=0.5, \gamma=3/2$, where we assume the radial distribution of density to be $\rho \propto r^{-n}$) 
  and the purple dashed line represents the ADAF solution ($n=3/2$, $\gamma=3/2$). 
  The normalization of the CDAF and ADAF are chosen arbitrarily.

  It is obvious that the simulated profiles roughly coincide with the CDAF solutions but only at middle angular ranges 
  ($20 - 50$ degrees). 
  The simulated density values are higher (or lower) than those of the CDAF at large (small) $\theta$ values. 
  The former indicates the presence of high density inflow, 
  whereas the latter is due to the outflow effect, which is not taken into account in the analytical solution.

\end{document}